\documentclass[preprint,aps]{revtex4}
\usepackage{graphicx}% Include figure files
\newcommand{\be}{\begin{equation}}
\newcommand{\ee}{\end{equation}}
\newcommand{\ba}{\begin{eqnarray}}
\newcommand{\ea}{\end{eqnarray}}
\newcommand{\p}{\partial}
\newcommand{\f}{\frac}

\begin{document}
\title
{Geometric Phase for a Nonstatic Coherent Light-Wave:
Nonlinear Evolution Harmonized with the Dynamical Phase
 \vspace{0.3cm}}
\author{Jeong Ryeol Choi\footnote{E-mail: choiardor@hanmail.net } \vspace{0.3cm}}

\affiliation{School of Electronic Engineering, Kyonggi University, %Gwanggyosan-ro,
Yeongtong-gu, Suwon,
Gyeonggi-do 16227, Republic of Korea \vspace{0.7cm}}

\begin{abstract}
\indent
Properties of the geometric phase for a nonstatic coherent light-wave
arisen in a static environment are analyzed from various angles.
The geometric phase varies in a regular nonlinear way, where the center of its
variation increases constantly with time.
This consequence is due to the effects of the periodic wave collapse and
expansion on the evolution of the geometric phase.
Harmonization of such a geometric-phase evolution with
the dynamical phase makes the total phase evolve with a unique
pattern that depends on the degree of nonstaticity.
The total phase exhibits a peculiar behavior for the case of
extreme nonstaticity, which is that it
precipitates periodically in its evolution, owing to a strong
response of the geometric phase to the wave nonstaticity.
It is confirmed that the geometric phase in the coherent state
is mostly more prominent compared to that in the Fock states.
For a simple case where the wave nonstaticity disappears, our
description of the geometric phase recovers
to the well-known conventional one which no longer undergoes periodical change.
While the familiar dynamical phase is just related to
the expectation value of the Hamiltonian, the geometric phase that we have managed
reflects a delicate nonstaticity difference in the evolution of quantum states.
\\
\\
%\indent \vspace{0.0cm} {\bf Keywords}: wave nonstaticity; geometric phase; coherent state; light wave;
%quantum state

\end{abstract}

\maketitle
\newpage

%\section{Introduction}
{\ \ \ } \\
{\bf %1. Introduction
1. INTRODUCTION
\vspace{0.2cm}} \\
Historically, the geometric phase was first recognized in 1931 by Dirac
in the form of a path dependent non-integrable phase \cite{pam,pam2}.
A great deal of attention has suddenly been paid to such an extra phase
after Berry's
seminal report \cite{ber,ber2} in 1984 in connection with
wave functions for dynamical systems with slowly varying parameters.
Soon after, Berry's development for a quantum phase has been extended to the case of nonadiabatic
wave evolutions by Aharonov and Anandan \cite{aha,aha2}.
Now, the concept of the geometric phase has been further extended to the case of non-cyclic and non-unitary evolutions of the
systems including light waves, leading to unveiling the profound
features of their quantum nature \cite{nmu,nmu2,nmu3}.

The geometric phase is a holonomy transformation that reveals %the
geometric character of the wave with a global phase change through
its evolution of modulo $2\pi$ \cite{pol,sfw}.
A slight different definition of the geometric phase is
the geometrical part of the phase evolution for
a wave in an arbitrary time $t$ (see, for example, Eq. (4.23) in Ref. \cite{mmc}).
We are interested in this latter definition of the geometric phase in this work,
because such a definition enables us to demonstrate nonlinear effects in the evolution of the
phase if any.

The geometric phase is noticeable since it directly affects
optical phenomena, such as quantum interference, quantum phase transitions,
parametric amplification, etc.
Furthermore, geometric phase can be applied to many next-generation
scientific technologies, including phase gates
for holonomic quantum computing \cite{aes,cvg}, detection and grating of a large-angle beam \cite{rpp},
and light control in holograms \cite{jke}.
In particular, geometric-phase gates in quantum computation allow us to process controllable
arithmetic logic operations \cite{cvg}.

If the parameters of a medium vary in time, a light wave in that medium may become nonstatic.
In a previous work \cite{nwh}, we have shown that nonstatic quantum light waves
can also appear even in a static environment.
We have investigated nonstatic quantum light waves and their associated phase properties
in the Fock states \cite{nwh,gow,pos}.
Subsequently, that research was
extended to the nonstatic waves in the coherent state \cite{ncs}.
Such waves exhibit a collapse and expansion in turn
in time as their characteristic of nonstaticity.
In those cases, a node and an antinode take place periodically in the graphic
of the time evolution of waves in the quadrature space.

Consequently, the eigenfunctions associated with the nonstatic waves are time-varying,
i.e., they are represented in terms of a time function
that obeys a nonlinear differential equation.
It was shown that such a variation of the eigenfunctions in the Fock states
leads to the appearance of a geometric phase
as the manifestation of peculiar wave nonstaticity \cite{gow}.
If we think that the geometric phase (latter definition) does not emerge in the ordinary Fock-state waves
described by the simple harmonic oscillator \cite{ze0,ze1,ze2,conn},
such a result in that work is worthy of note.

In this work, we will investigate the geometric phase for a nonstatic light
wave arisen in a static environment through the same sprit of Ref. \cite{gow},
but in the coherent state \cite{gis0,gis}.
The importance of coherent states is that it is
possible to manage the correlation and coherence properties of light waves
by means of them.
The preservation and enhancement of quantum coherence is needed
in fulfilling quantum algorithms in quantum computation with qubits.
We elucidate the relation between nonstatic-wave mechanism and its resultant geometric phase
in the coherent state through this research.
To get a keen
insight for the kinematics of the phase evolution with nonstaticity,
the clarification of the behavior of
accompanying geometric phases is indispensable.

The effects of the periodic wave collapse and expansion on the
evolution of the considered geometric phase will be analyzed.
Similarities and differences of our consequence for the phase evolution
in the coherence state and that in the Fock states will be shown.
The geometric phase will also be compared to the ordinary dynamical phase and we clarify
their interplay and concurrent relation with harmony
for the formation of a novel phase outcome.
Phase properties with an extreme nonstaticity will be additionally seen.
Through these,
we deepen the understanding for the characteristics of the quantum phase
in general nonstatic coherent waves and
demonstrate the geometrical features of the wave evolution.
\\
\\
%\section{Description of the nonstatic coherent state}
{\bf %2. Description of the Nonstatic Coherent State
2. DESCRIPTION OF THE NONSTATIC COHERENT STATE
\vspace{0.2cm}} \\
Let us consider a nonstatic light
wave in a non-dissipative medium where the electric permittivity
$\epsilon$ and magnetic permeability $\mu$ are independent of time.
To investigate the geometric phase strictly, we need to develop a theory
for the phenomena of quantum coherent wave with nonstaticity together with
the structure of the related quantum state.
The Hamiltonian which describes light-wave evolution in the medium is
given by
\be
\hat{H}={\hat{p}^2}/{(2\epsilon)} + \epsilon\omega^2 \hat{q}^2/2 , \label{1}
\ee
where $\hat{q}$ is the operator representation of the canonical variable
$q(t)$ that gives amplitude of the wave and $\hat{p} = -i\hbar \p /\p q$.
$q(t)$ in this setup corresponds to the time function in the vector potential
when we separate it into time and position functions in the Coulomb gauge \cite{nwh,ncs}.
The angular frequency $\omega$ of the wave satisfies $\omega =k c$ where
$k$ is the wave number, whereas the velocity of the wave is expressed as
$c=1/\sqrt{\epsilon\mu}$.
If we denote a general classical solution for the variable $q$ as $Q_{\rm cl}(t)$,
it can be represented in the form
\be
Q_{\rm cl}(t) = Q_0 \cos\tilde{\theta}(t),  \label{}
\ee
where $Q_0$ is a constant, $\tilde{\theta}(t)=\omega (t-t_0) +\theta_0$,
$t_0$ is an initial time, and $\theta_0$ is an arbitrary phase at $t_0$.
We consider only the case where $t \geq t_0$ for convenience throughout this work.

A nonstatic wave is described by a time function of the form \cite{nwh,ncs}
\be
f(t) = c_1 \sin^2 \tilde{\varphi}(t)+ c_2 \cos^2 \tilde{\varphi}(t) +
c_3 \sin [2\tilde{\varphi}(t)], \label{4}
\ee
where $\tilde{\varphi}(t)=\omega (t-t_0) +\varphi$ and $\varphi$ is a phase at $t=t_0$,
while $c_1$, $c_2$, and $c_3$ are real constants which give the nonstaticity and they
satisfy the relation
\be
c_1c_2-c_3^2 = 1, \label{5}
\ee
with
\be
c_1c_2 \geq 1. \label{c5+1}
\ee
A generalized
annihilation operator associated with the nonstatic light is represented as \cite{ncs}
\be
\hat{A}= \sqrt{\f{\epsilon\omega}{2\hbar f(t)}}\left(1- i \f{\dot{f}(t)}{2\omega}\right) \hat{q} + i\sqrt{\f{
f(t)}{2\epsilon\omega\hbar}} \hat{p}.  \label{7}
\ee
The Hermitian adjoint of this operator $\hat{A}^\dagger$
is the creation operator, whereas the two operators satisfy the boson commutation relation
such that $[\hat{A}, \hat{A}^\dagger]=1$.
We write the eigenvalue equation for $\hat{A}$ in the form
\be
\hat{A} |A \rangle =A |A \rangle, \label{8}
\ee
where $A$ is the eigenvalue and $|A \rangle$ is the eigenfunction.
Then, $|A \rangle$ is a generalization of the coherent state considering nonstaticity.
If we set
\be
\hat{I}= \hbar \omega (\hat{A}^\dagger \hat{A}+ 1/2), \label{ioM}
\ee
$\hat{I}$ is an
invariant operator \cite{ncs}. While the Hamiltonian itself ($\hat{H}$) is also an invariant operator,
$\hat{I}$ is regarded as a generalized invariant operator and is reduced to $\hat{H}$ when
$c_1=c_2=1$ and $c_3=0$.

Before we enter the nonstatic coherent state,
let us look into the nonstatic Fock states as a preliminary step.
The Fock-state wave functions with nonstaticity are given by \cite{nwh}
\be
\langle q |\Psi_n(t) \rangle = \langle q |\Phi_n(t) \rangle \exp [i\gamma_n(t) ]~~~~~~~n=0,1,2,\cdots, \label{M32}
\ee
where the associated eigenfunctions $\langle q |\Phi_n(t) \rangle$ and phases $\gamma_n(t)$ are
expressed as
\ba
\langle q |\Phi_n(t) \rangle &=& \left({\f{\zeta(t)}{\pi}}\right)^{1/4} \f{1}{\sqrt{2^n
n!}} H_n \left( \sqrt{\zeta(t)} q \right) \exp \left[
- \f{1}{2}\zeta(t) \left(1-i\f{\dot{f}(t)}{2\omega}\right) q^2 \right], \label{M33} \\
\gamma_n(t) &=& {-\omega (n+1/2) \int_{t_0}^t f^{-1} (t') dt'} + \gamma_n (t_0),
 \label{M34}
\ea
while $H_n$ are $n$th order Hermite polynomials and $\zeta(t) = {\epsilon\omega}/{[\hbar f(t)]}$.
All the states which retain their functional form during nonstatic evolution can be
represented as a superposition of $\langle q |\Psi_n(t) \rangle$ such that
\be
\langle q |\Psi(t) \rangle = \sum_{n=0}^\infty a_n \langle q |\Psi_n(t) \rangle ,  \label{qA-1}
\ee
where $a_n$ are complex numbers that obey $\sum_{n=0}^\infty |a_n|^2 =1$.
Interference between the phases of different components in Eq. (\ref{qA-1})
is responsible for the emergence of nonclassical effects in the resultant
state, such as oscillation in photon-number distribution \cite{opn},
quadrature squeezing \cite{sns}, and sub-Poissonian photon statistics \cite{fls}.
The generalized
coherent state $\langle q |A \rangle$ that we are interested in this work
can also be represented in a similar way. Namely \cite{lou},
\be
\langle q |A \rangle  = \sum_{n=0}^\infty b_n (A) \langle q |\Phi_n \rangle ,  \label{qA}
\ee
where the transformation functions are given by
$b_n (A) = \exp{(-|A|^2/2 )} {A^n}/{\sqrt{n!}}$.
Although we have represented Eq. (\ref{qA}) in a slightly different manner compared to Eq. (\ref{qA-1}),
the full wave function in the coherent state, which includes its phase,
can also be represented as Eq. (\ref{qA-1}), i.e., can be expressed as an expansion
in terms of $\langle q |\Psi_n(t) \rangle$ (see Appendix A).

The coherent state that has nonstatic characteristic
was derived using Eq. (\ref{8}) in our previous work which is Ref. \cite{ncs}.
The resultant eigenfunction in the configuration space is of the form
\be
\langle q |A \rangle = \sqrt[4]{\f{\zeta(t)}{\pi }}
\exp \Bigg[ -\f{\zeta(t)}{2} \bigg( 1-
i \f{\dot{f}(t)}{2\omega} \bigg)q^2 + \sqrt{{2\zeta(t)}} A q
-\f{1}{2} |A|^2-\f{1}{2} A^2 \Bigg]. \label{19}
\ee
Though this state exhibits nonstaticity, it is overcomplete
and non-orthogonal as usual coherent states.
On the other hand, the eigenvalue associated to this state is given by \cite{ncs}
\be
A(t) = A_0 e^{-i[\omega T(t)+\theta]}, \label{18}
\ee
where $A_0$ is its amplitude, which is of the form
\be
A_0 = \Bigg\{ \f{\epsilon\omega}{2\hbar}\Bigg[\f{\cos^2 \tilde{\theta}(t)}{f(t)}+
\Bigg( \f{\dot{f}(t)}{2\omega\sqrt{f(t)}}\cos \tilde{\theta}(t)
+\sqrt{f(t)}\sin \tilde{\theta}(t) \Bigg)^2 \Bigg] \Bigg\}^{1/2}Q_0 , \label{EA0}
\ee
while $\theta$ is the phase of $A(t)$ at $t_0$, and
\be
T(t) = G(t) - G(t_0) +{\mathcal G}(t)/\omega, \label{28} \\
\ee
with
\be
G(\tau) = \f{1}{\omega}\tan^{-1} \{ c_3+c_1\tan[\omega (\tau-t_0) +\varphi] \}, \label{29}
\ee
\be
{\mathcal G}(t)=\pi \sum_{m=0}^{\infty}u[t-t_0-(2m+1)\pi/(2\omega)+\varphi/\omega],
\label{gts}
\ee
whereas $u[x]$ is the Heaviside step function.
We note that $A_0$ is a time constant, but dependent on $c_1$, $c_2$,
$\varphi$, and $\theta_0$.
In the expression of Eq. (\ref{28}) with Eqs. (\ref{29}) and (\ref{gts}),
we have restricted the region of $\varphi$ as $-\pi/2 \leq \varphi < \pi/2$
for convenience without loss of generality.
This interval is enough because the geometric phase varies with the period
$\pi$ for $\varphi$.
In fact, $T(t)$ can also be represented as \cite{ncs}
\be
T(t) = \int_{t_0}^t f^{-1}(t')dt'. \label{Tt}
\ee

%%%%%%%%%%%%%%%%%%%%%%%%%%%%%%%%%%%%%%
\begin{figure}%[H]
\centering
\includegraphics[keepaspectratio=true]{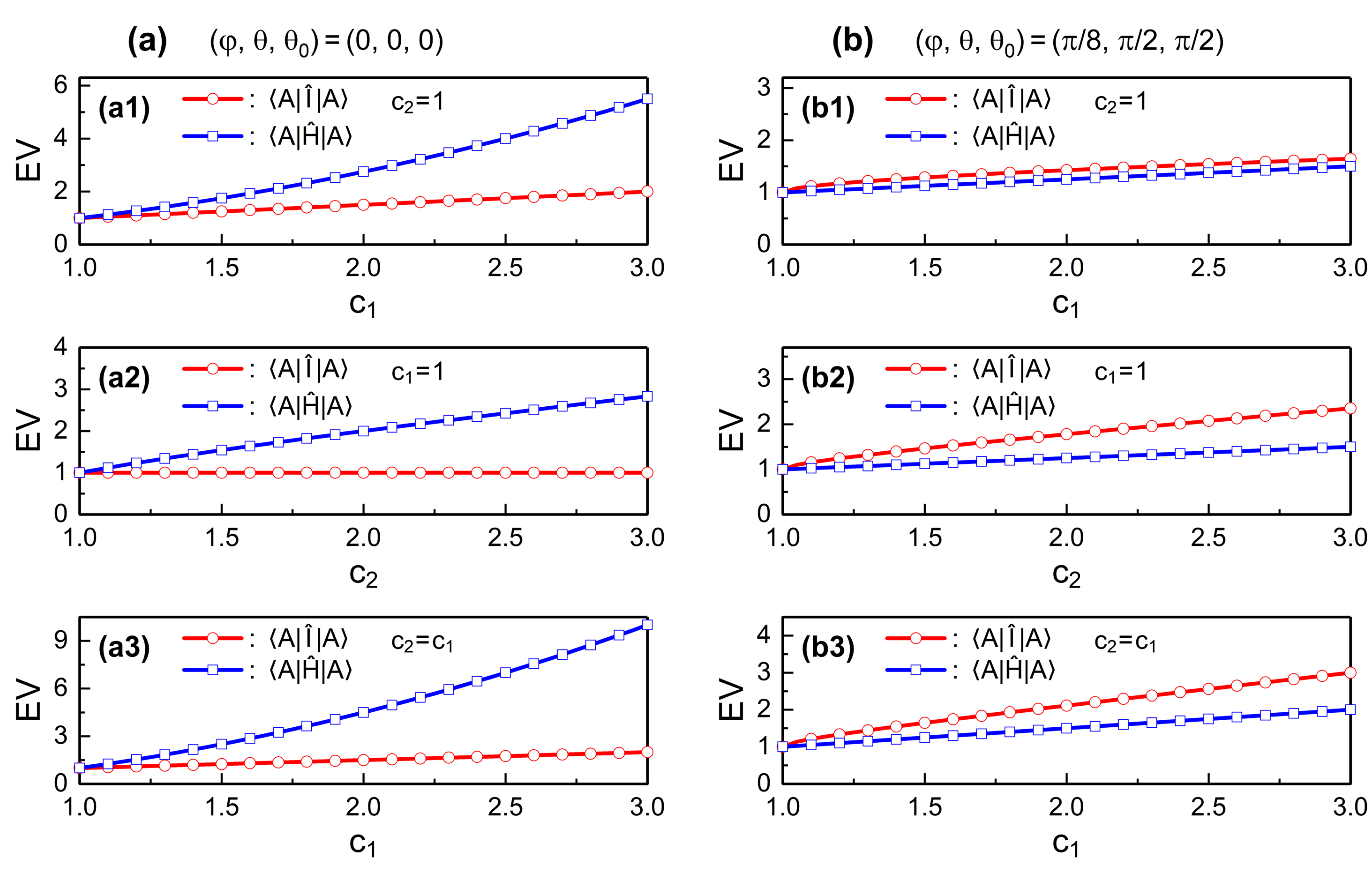}
\caption{\label{Fig1} ({\bf a}) is the relation of the expectation
value (EV) of $\hat{I}$ ($\langle A|\hat{I}|A \rangle$) and
$\hat{H}$ ($\langle A|\hat{H}|A \rangle$) with $c_1$ ({\bf a1}, {\bf a3}) and $c_2$ ({\bf a2})
in the coherent state, where angles are chosen as
($\varphi$, $\theta$, $\theta_0$)$=$(0, 0, 0): we have taken
$c_2=1$ for ({\bf a1}), $c_1=1$ for ({\bf a2}), and $c_2=c_1$ for ({\bf a3}).
({\bf b}) is the same as ({\bf a}), but for the case of a different angles choice which is
($\varphi$, $\theta$, $\theta_0$)$=$($\pi/8$, $\pi/2$, $\pi/2$).
$c_3$ is determined based on the values of $c_1$ and $c_2$ according to Eq. (\ref{5}).
Among two possible values of $c_3$ which are $c_3= \pm \sqrt{c_1c_2-1}$,
we have chosen $c_3= \sqrt{c_1c_2-1}$ in this figure for convenience: this convention
will also be used for all subsequent figures.
We have used $t=1$, $t_0=0$, $Q_0=1$, $\omega=1$, $\hbar=1$, and $\epsilon=1$.
The EV $\langle A|\hat{I}|A \rangle = \hbar \omega (A_0^2+ 1/2)$,
depicted here, is evaluated based on
Eq. (\ref{ioM}) with the eigenvalue relation Eq. (\ref{8}),
utilizing the formula of $A(t)$ given in Eq. (\ref{18}) and its complex conjugate $A^*(t)$
which are described with Eqs. (\ref{EA0})-(\ref{gts}).
The formula of $\langle A|\hat{H}|A \rangle$ is also evaluated with the same procedure
using Eq. (\ref{1}), after converting ($\hat{q}$, $\hat{p}$) in the Hamiltonian
into the expression with ($\hat{A}$, $\hat{A}^\dagger$)
based on Eq. (\ref{7}) and its Hermitian-conjugate representation.
 }
\end{figure}
%%%%%%%%%%%%%%%%%%%%%%%%%%%%%%%%%%%%%%

The magnitude of nonstaticity for a wave can be estimated by a measure of nonstaticity.
For the coherent state, Eq. (\ref{19}), such a measure is determined by $c_1$ and $c_2$ such that \cite{nwh,ncs}
\be
D = \f{\sqrt{(c_1+c_2)^2-4}}{2\sqrt{2}}.  \label{Mn}
\ee
This is the same as the nonstaticity measure in the Fock states.

The coherent state described up until now can be applied to
the analysis of the system.
For instance, let us see the behavior of $\hat{I}$ and $\hat{H}$
in the coherent state.
Though $\hat{I}$ and $\hat{H}$ do not vary over time, we see from Fig. 1 that
their expectation values are dependent on $c_1$ and $c_2$.
Figure 1({\bf a}) is the case where $\langle A|\hat{I}|A \rangle$
is smaller than $\langle A|\hat{H}|A \rangle$, whereas Fig. 1({\bf b})
the case where $\langle A|\hat{I}|A \rangle$
is larger than $\langle A|\hat{H}|A \rangle$.
The difference of $\langle A|\hat{I}|A \rangle$ from
$\langle A|\hat{H}|A \rangle$ becomes large as $c_1$ and/or $c_2$ increase in most cases,
but $\langle A|\hat{I}|A \rangle$ is reduced to $\langle A|\hat{H}|A \rangle$
in the limit $c_1=c_2 \rightarrow 1$.

The wave description developed until now in the coherent state
is also necessary in evaluating phases of the nonstatic light,
because the characteristics of the geometric phase are determined
depending on the specific formula of wave functions.
The geometric phase of the nonstatic quantum light wave will be investigated
in the following sections using that description.
\\
\\
%\section{Geometric phase}
{\bf %3. Generalized Nonlinear Geometric Phase
3. GENERALIZED NONLINEAR GEOMETRIC PHASE
\vspace{0.2cm}} \\
{\bf 3.1. Integral Representation}
\\
We focus on the geometric phase that takes place due to nonstaticity
in the wave evolutions of quantum light.
Several mathematical methods for formulating geometric phases are known. Some of them are
the methods utilizing the Schr\"{o}dinger equation \cite{pol,hka0}, a path integral technique \cite{hka,hka1,hka2},
and the canonical transformation approach \cite{snb,snb2}.
The former method (the method based on the Schr\"{o}dinger equation)
is much more general than others and usually adopted
in the relevant studies owing to its simplicity.
We will also use it in the present work.

In Ref. \cite{gow}, we have studied the properties of the geometric phases for nonstatic waves in
the Fock states as mentioned previously.
The geometric phase for a superposition of Fock states, including coherent states, can also
be derived in the same way adopted in that report.
Hence, it is possible to extend the method given in Ref. \cite{gow}
to our coherent state through the expression given in Eq. (\ref{qA}).
However, since the wave function in the coherent state is known in Eq. (\ref{19})
at this stage, the use of that wave function instead of Eq. (\ref{qA})
is more convenient for the derivation of the associated geometric phase.

Bearing in mind this, let us evaluate the geometric phase of the nonstatic light,
in addition to the usual dynamical phase, in the coherent state.
The definitions of the geometric and dynamical phases are given by
\ba
\gamma_{G}(t) &=&  \int_{t_0}^t \langle A(t')
|i\frac{\partial}{\partial t'}| A(t') \rangle dt' +\gamma_{G}(t_0),
\label{20} \\
\gamma_{D}(t) &=& - \f{1}{\hbar} \int_{t_0}^t \langle A(t') |
\hat{H}(\hat{q},\hat{p},t')| A(t') \rangle dt' + \gamma_{D}(t_0). \label{21}
\ea
If we think that Eq. (\ref{20}) includes a time derivative of the eigenfunction,
the geometric phase occurs only when the eigenfunction is described in terms of
time \cite{conn}.
The geometric phase defined in Eq. (\ref{20}) is gauge-invariant:
an exemplary proof of this characteristic is shown in Appendix B.
Because the nonstatic eigenfunction given in Eq. (\ref{19}) is dependent on time,
the geometric phase in this coherent-state description is non-zero.
The time derivative of the eigenfunction, Eq. (\ref{19}), in the configuration space yields
\ba
\f{\p \langle q |A \rangle}{\p t} &=& \Bigg[\f{i\omega}{f(t)}A^2-\f{\dot{f}(t)}{4f(t)}
-\sqrt{2\zeta(t)}\f{A}{f(t)}\bigg( \f{\dot{f}(t)}{2} + i\omega \bigg)q \nonumber \\
& &+\bigg( \f{i\epsilon}{8\hbar f^2(t)}[2\omega -i\dot{f}(t)]^2 -\f{i\epsilon\omega^2}{2\hbar} \bigg)q^2
\Bigg]\langle q |A \rangle. \label{22}
\ea
A rigorous evaluation of Eq. (\ref{20}) after inserting this formula gives
\ba
\gamma_{G}(t) &=& \int_{t_0}^t \Gamma_G(t')  dt' +\gamma_{G}(t_0),
\label{23}
\ea
where
\be
\Gamma_G(t) =  \f{\omega}{4 }g_1(t) +\f{1}{16\omega }g_2(t) +\f{1}{4}g_3(t)
+\f{\omega }{4}g_4(t) , \label{GGt}
\ee
while $g_i(t)$ ($i=$1--4) are time functions that are represented as (see Appendix C)
\ba
g_1(t)&=& -\f{1}{ f(t)}\{2A_0^2\cos [2(\omega T(t)+\theta)]-2A_0^2+1\}, \label{24}  \\
g_2(t)&=&\f{[\dot{f}(t)]^2}{f(t)}\{2A_0^2\cos [2(\omega T(t)+\theta)]+2A_0^2+1\}, \label{25}  \\
g_3(t)&=&-\f{2\dot{f}(t)}{ f(t)}A_0^2\sin [2(\omega T(t)+\theta)], \label{26}  \\
g_4(t)&=& f(t)\{2A_0^2\cos [2(\omega T(t)+\theta)]+2A_0^2+1\}. \label{27}
\ea

For the non-displaced case where $A_0=0$, Eq. (\ref{23}) is reduced to
\ba
\gamma_{G}(t) &=& \f{1}{4} \int_{t_0}^t
\bigg( \omega f(t') - \f{\omega}{f(t')} +\f{[\dot{f}(t')]^2}{4\omega f(t')} \bigg)
 dt' +\gamma_{G}(t_0).
\label{23nd}
\ea
This is the same as the geometric phase in the Fock states
with $n=0$ (see Eqs. (9)-(13) in Ref. \cite{gow}).
This correspondence is natural because the coherent state
is the displaced one of the zero-point Fock state.
\\
\\
{\bf 3.2. Analytical Formula of Geometric Phase}
\\
{\it 3.2.1. Deriving geometric phase}
\\
Direct acquisition of the geometric phase
from Eq. (\ref{23}) may be not easy because it involves a complicated integration.
Our strategy for overcoming this difficulty is that we turn our attention
to the dynamical phase for the moment, which seems less hard, and therein we completely derive it.
Then we represent the geometric phase in terms of the factor(s)
used in the dynamical phase, expecting that this manipulation
leads to full representation of the geometric phase in a reasonable way.
To obtain the dynamical phase, we insert the Hamiltonian Eq. (\ref{1}) into Eq. (\ref{21})
after expressing the canonical variables
in terms of the ladder operators $\hat{A}$ and $\hat{A}^\dagger$. Then, using Eqs. (\ref{8}) and (\ref{18}), we get
\ba
\gamma_{D}(t) &=& \int_{t_0}^t \Gamma_D(t')  dt' +\gamma_{D}(t_0),
\label{30}
\ea
where
\be
\Gamma_D(t) = - \bigg( \f{\omega}{4 }\bar{g}_1(t) +\f{1}{16\omega }g_2(t) +\f{1}{4}g_3(t)
+\f{\omega }{4}g_4(t) \bigg),  \label{D30}
\ee
while (see Appendix C)
\be
\bar{g}_1(t)= -\f{1}{ f(t)}\{2A_0^2\cos [2(\omega T(t)+\theta)]-2A_0^2-1\}. \label{31}
\ee
As can be seen from Appendix D, $\Gamma_D(t)$ is a time constant though it is expressed
in terms of time.
Because Eq. (\ref{21}) means that $\Gamma_D(t)$ is directly related to the expectation value of
$\hat{H}$, this consequence is not strange and closely related to
the energy conservation law.
Since $\Gamma_D(t)$ does not vary over time, we can write
Eq. (\ref{30}) without the integral symbol in the form
\ba
\gamma_{D}(t) &=& \Gamma_D(\bar{t}) [t-t_0] +\gamma_{D}(t_0),
\label{gD}
\ea
where $\bar{t}$ is an arbitrary fixed time that satisfies $\bar{t} \geq t_0$.
The time fixation $t = \bar{t}$ in $\Gamma_D$ is not a necessary condition in
the above equation, but the related mathematics becomes simpler by treating Eq. (\ref{gD}) with such a fixed time.

Now, in the case of the geometric phase, we take attention to the fact that
Eq. (\ref{GGt}) can be rearranged in a way that it involves $\Gamma_D(t)$ such that
\be
\Gamma_{G}(t) = -\f{\omega}{2f(t)} - \Gamma_D(t) .
\label{gGa}
\ee
Hence, employing again the fact that $\Gamma_D(t)$ is constant over time
together with the relation in Eq. (\ref{Tt}),
the geometric phase described in terms of Eq. (\ref{gGa})
can also be reduced to a simple form as
\be
\gamma_{G}(t) = -\f{1}{2}\omega T(t) - \Gamma_D(\bar{t}) [t-t_0] +\gamma_{G}(t_0).
\label{gG}
\ee
Because we know not only the formula of $T(t)$ in the above equation from
Eq. (\ref{28}) with Eqs. (\ref{29}) and (\ref{gts}), but the formula of $\Gamma_D(\bar{t})$
from Eq. (\ref{D30}) with Eqs. (\ref{31}), (\ref{25}), (\ref{26}), and (\ref{27})
as well, we now have identified the complete analytical formula of the geometric
phase in the coherent state.
This formula is more general than the existing ones \cite{snb,5} concerning the coherent state.
We note that the geometric phase appeared in the previous reports (Refs. \cite{snb,5}) is the one that
does not considered the wave nonstaticity.
The geometric phase given in Eq. (\ref{gG})
exhibits nonlinear effects due to the wave nonstaticity.
By the way, it is impossible to gauge out the geometric part in phase because of
the fact that the geometric phase is gauge invariant.

If we take $c_1=c_2=1$ and $c_3=0$, the nonlinear effects in the geometric phase
vanish, leading to the geometric phase being a reduced familiar one:
\be
\gamma_{G}(t) =  \omega A_0^2 (t-t_0) +\gamma_{G}(t_0), \label{27+1}
\ee
which coincides with the formula given in Ref. \cite{igp}.
If the displacement of the quadrature oscillation also
disappears ($A_0 \rightarrow 0$) along with this, the  geometric phase no longer exists.
We can thereby conclude that there are two sources of the appearing of the geometric phase
for the present case. One is the wave nonstaticity and the other is the displacement of the oscillation.
On the other hand, the case of Fock states treated in Refs. \cite{nwh,gow},
the origin of the geometric phase is only the wave nonstaticity.
%%%%%%%%%%%%%%%%%%%%%%%%%%%%%%%%%%%%%%
\begin{figure}%[H]
\centering
\includegraphics[keepaspectratio=true]{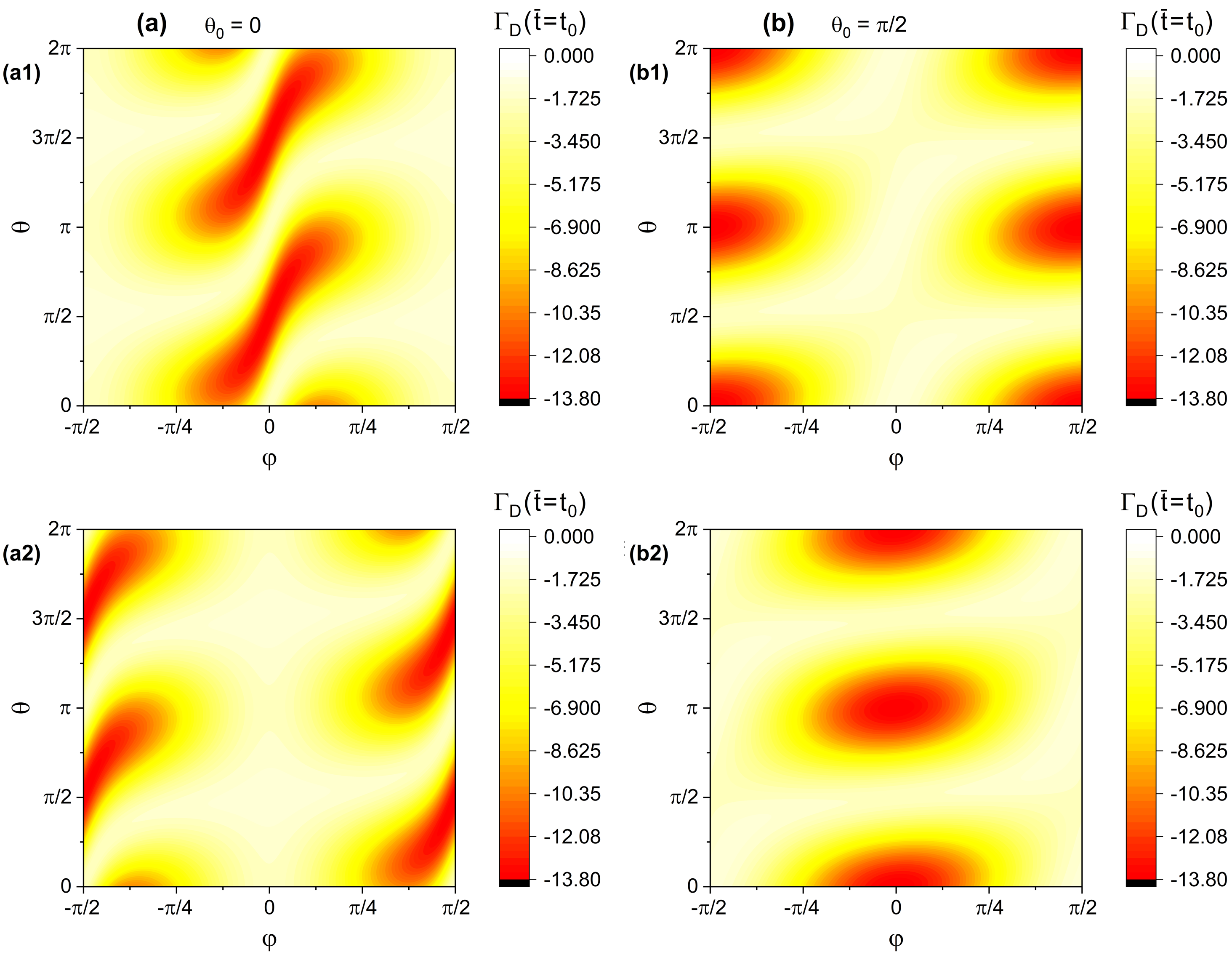}
\caption{\label{Fig2} Density plots of $\Gamma_{D}(\bar{t}=t_0)$
as a function of $\varphi$ and $\theta$,
where Eq. (\ref{Dt0}) with Eqs. (\ref{EA0}), (\ref{k1}), and (\ref{k2})
is used as its formula.
We have chosen $\theta_0=0$  for ({\bf a}) and $\theta_0=\pi/2$ for ({\bf b}),
whereas ($c_1$, $c_2$) are taken as (5.0, 0.2) for subpanels ({\bf a1}) and ({\bf b1}),
and as (0.2, 5.0) for ({\bf a2}) and ({\bf b2}).
We have used $Q_0=1$, $\omega=1$, $\epsilon=1$, and $\hbar=1$.
 }
\end{figure}
%%%%%%%%%%%%%%%%%%%%%%%%%%%%%%%%%%%%%%
\\
{\it 3.2.2. Characteristics of $\Gamma_{D}(\bar{t})$}
\\
Because the geometric phase is represented in terms of $\Gamma_{D}(\bar{t})$,
it is important to elucidate the characteristics of $\Gamma_{D}(\bar{t})$.
If we take $\bar{t}=t_0$ without loss of generality
under the condition given in Eq. (\ref{5}), $\Gamma_{D}(\bar{t})$ becomes
\ba
\Gamma_{D}(\bar{t}=t_0) &=& - \f{\omega}{16 \kappa_1}[4(1+4A_0^2\sin^2\theta)+
(1+4A_0^2\cos^2\theta)(4\kappa_1^2 + \kappa_2^2) \label{Dt0}
\nonumber \\
& &-8A_0^2\sin(2\theta)\kappa_2],
\ea
where
\ba
\kappa_1 &=& c_1 \sin^2 \varphi +c_2 \cos^2 \varphi + \sqrt{c_1c_2-1}\sin(2\varphi),
\label{k1} \\
\kappa_2 &=& (c_1-c_2) \sin (2 \varphi) + 2\sqrt{c_1c_2-1}\cos(2\varphi).
\label{k2}
\ea
$\Gamma_{D}(\bar{t})$ in Eq. (\ref{Dt0}) is universally valid since
the choice of $\bar{t}$ does not affect the result.
Figure 2 represents the relation of $\Gamma_{D}(\bar{t}=t_0)$
on angles $\varphi$, $\theta$, and $\theta_0$.
We confirm from each panel of this figure
that $\Gamma_{D}(\bar{t}=t_0)$ is different depending on
$\varphi$ and $\theta$ in a regular manner; moreover, the comparison between subpanels ({\bf a1})
and ({\bf b1}) (or ({\bf a2}) and ({\bf b2}))
shows that $\Gamma_{D}(\bar{t}=t_0)$ also depends on $\theta_0$.
Additional comparison between ({\bf a1}) and ({\bf a2}) (or ({\bf b1})
and ({\bf b2})) shows the difference of $\Gamma_{D}(\bar{t}=t_0)$ depending on
$c_1$ and $c_2$.
Meanwhile, the magnitude of $\Gamma_{D}(\bar{t})$ can be
regulated at our will by adjusting the value of $Q_0$ (or $A_0$ instead).
This regulation does not affect the measure of nonstaticity since the measure
is determined only by $c_1$ and $c_2$ as can be confirmed from Eq. (\ref{Mn}).

If we further take $\varphi=\theta=0$, Eq. (\ref{Dt0}) is simplified to
\be
\Gamma_{D}(\bar{t}=t_0) =-\omega \bigg[\f{c_1+c_2}{4}
+  A_0^2 \bigg(c_1+c_2 -\f{1}{c_2}\bigg)\bigg]. \label{vt0}
\ee
For the static case ($c_1=c_2=1$), this becomes
\be
\Gamma_{D}(\bar{t}=t_0) =-\omega \bigg( A_0^2 + \f{1}{2} \bigg). \label{gDntr-}
\ee
By the way, for the comparison purpose, we have represented the consequence of wave phases
in the Fock states in Appendix E.
The formula in Eq. (\ref{gDntr-}) is similar to the following familiar form in the Fock states,
which can be obtained from Eq. (\ref{gDnt}) in Appendix E with the same choice of
the nonstaticity parameters, $c_1=c_2=1$:
\be
\f{\gamma_{D,n}(t)-\gamma_{D,n}(t_0)}{t-t_0} = -\omega \left( n+\f{1}{2} \right) . \label{gDntr}
\ee
It looks like that $A_0^2$ in Eq. (\ref{gDntr-}) plays the same role
as $n$ in Eq. (\ref{gDntr}). This fact will be used in the later investigation.
\\
\\
{\bf 3.3. Total Phase and Nonlinearity Perspectives}
\\
The total phase acquired by the state in the course of its
evolution is the sum of the geometric and dynamical phases:
\be
\gamma(t) = \gamma_{G}(t)+\gamma_{D}(t) = -\f{1}{2}\omega T(t) + \gamma(t_0), \label{32}
\ee
where $\gamma(t_0)$ is the phase at $t_0$.
This is the same as the minimum evolution of the overall phase for the light
in the Fock states, i.e., the value of Eq. (\ref{32}) is identical to
that in the Fock states with the zero-point quantum number ($n=0$),
which appears in Eq. (\ref{gnt}) in Appendix E.

Regarding the phase shown
in Eq. (\ref{32}), the wave function in the coherent state is given by
\be
\langle q | \Psi_{\rm coh}(t) \rangle = \langle q | A \rangle e^{i\gamma (t)}. \label{32w}
\ee
It may be possible to investigate the complete nonstatic effects of the displaced
wave utilizing this generalized wave function from the theoretical point of view.

By the way, it may be instructive to represent $\gamma_{G}(t)$ and $\gamma(t)$ in the forms
\ba
\gamma_{G}(t) &=& \gamma_{G, {\rm NL}}(t) + \gamma_{G, {\rm L}}(t),
\label{gnl}
\\
\gamma(t) &=& \gamma_{\rm NL}(t) + \gamma_{\rm L}(t), \label{gnl-}
\ea
where $\gamma_{G, {\rm NL}}(t)$ and $\gamma_{\rm NL}(t)$ are
periodical nonlinear terms whereas
$\gamma_{G, {\rm L}}(t)$ and $\gamma_{\rm L}(t)$ are linear terms, which are given by
\ba
\gamma_{G, {\rm NL}}(t) &=& \gamma_{\rm NL}(t) = -\f{1}{2}\omega [T(t)-(t-t_0)]
+ \gamma_{G, {\rm NL}}(t_0), \label{ant1} \\
\gamma_{G, {\rm L}}(t) &=& - \bigg(\Gamma_D(\bar{t})+ \f{1}{2}\omega \bigg) (t-t_0)
+ \gamma_{G, {\rm L}}(t_0), \label{ant2} \\
\gamma_{\rm L}(t) &=& - \f{1}{2}\omega (t-t_0)
+ \gamma_{\rm L}(t_0), \label{ant3}
\ea
together with the trivial relation $\gamma_{\rm NL}(t_0) = \gamma_{G, {\rm NL}}(t_0)$.
Nonlinear effects taken place by the nonlinear terms given above
are apparently a matter of concern and interest.
Nonlinearity in the geometric phase and its
consequence in relation with the total phase will be investigated in detail in Sec. 5.
\\
\\
{\bf %4. Validity of the General Geometric Phase
4. COMPLETENESS OF THE GENERAL GEOMETRIC PHASE
\vspace{0.2cm}} \\
The geometric phase given in Eq. (\ref{gG}) (or its non-integrated formula in Eq. (\ref{23}))
is a generalized one regarding wave
nonstaticity of the coherent state, whereas it is reduced to the well-known
standard one when $c_1=c_2 \rightarrow 1$.
We will now see the validity and completeness of such a geometric phase
by examining whether the full wave function developed considering it
obeys the Schr\"{o}dinger equation or not.
By taking time derivative of the wave function, Eq. (\ref{32w}), after
inserting Eqs. (\ref{19}) and (\ref{32}) with Eq. (\ref{18}) in it, we obtain
\ba
\f{\p \langle q|\Psi_{\rm coh}(t)\rangle}{\p t}&=& \f{e^{V(t)}}{4\hbar f^2(t)}
\sqrt[4]{\f{\zeta(t)}{\pi}} [ W(t) q^2 - X(t)q -Y(t)] \label{mbt},
\ea
where
\ba
V(t)&=&\sqrt{2\zeta(t)} A_0 e^{-i(\omega T+\theta)} q-
\{ A_0^2(1+ e^{-2i(\omega T+\theta) }) +4i\theta +5i\omega T  \nonumber \\
& &+\zeta(t) [1-i\dot{f}(t) /(2\omega)] q^2 \}/2, \\
W(t) &=& \{2 \omega \dot{f}(t) -i [\dot{f}(t)]^2
+if(t) \ddot{f}(t) \}e^{2i(\omega T+\theta)} \epsilon, \\
X(t)&=& 2\sqrt{2\zeta(t)} \hbar  A_0 e^{i(\omega T+\theta)}[ f(t) \dot{f}(t)
+ 2i \omega f^2(t)\dot{T} ], \\
Y(t)&=& e^{2i(\omega T+\theta)}\hbar f(t) \dot{f}(t)
+ 2i\hbar \omega f^2(t)\dot{T}[e^{2i(\omega T+\theta)} -2A_0^2].
\ea
We used the relation $d A_0/d t =0$ in this evaluation, but did/need not
resort to the formula of $A_0$ given in Eq. (\ref{EA0}) in this case.
On the other hand, the second order differentiation of the wave
function with respect to $q$ is of the form
\ba
\f{\p^2 \langle q|\Psi_{\rm coh}(t)\rangle}{\p q^2}&=& \sqrt[4]{\f{\zeta(t)}{\pi }} e^{-Z(t)/2}
\{ \{\sqrt{2\zeta(t)} A_0 e^{-i(\omega T+\theta) }
-\zeta(t)[1-i\dot{f}(t)/(2\omega)] q \}^2
\nonumber \\
& &-\zeta(t) [1-i\dot{f}(t) /(2\omega)] \}, \label{mbq2}
\ea
where
\be
Z(t)= A_0^2(1+ e^{-2i(\omega T+\theta) }) -2\sqrt{2\zeta(t)} A_0 e^{-i(\omega T+\theta) } q
+i\omega T+\zeta(t) [1-i\dot{f}(t) /(2\omega)] q^2 .
\ee
Now we regard the following relations:
\ba
\dot{T} &=&1/f(t),  \label{dtt} \\
\ddot{f}(t) &=&\f{[\dot{f}(t)]^2}{2f(t)}-2\omega^2 [f(t)-1/f(t)]. \label{dft}
\ea
The relation in Eq. (\ref{dtt}) can be recognized from Eq. (\ref{Tt}).
Another relation, Eq. (\ref{dft}), is shown in Eq. (3) of Ref. \cite{ncs}.
By utilizing Eqs. (\ref{mbt}) and (\ref{mbq2}) after inserting
Eqs. (\ref{dtt}) and (\ref{dft}) into Eq. (\ref{mbt}),
it is possible to confirm that the wave function satisfies the
Schr\"{o}dinger equation such that
\be
i\hbar \f{\p \langle q|\Psi_{\rm coh}(t)\rangle}{\p t} =-\f{\hbar^2}{2\epsilon} \f{\p^2 \langle q|\Psi_{\rm coh}(t)\rangle}{\p q^2}
+\f{1}{2} \epsilon\omega^2 q^2 \langle q|\Psi_{\rm coh}(t)\rangle. \label{q46}
\ee
Thus, the total phase that we have considered is correct.
This result
also means that its components, the geometric phase (and the dynamical phase),
are exactly evaluated, demonstrating that our geometric phase
for the nonstatic wave is right.
We note that the wave functions in Fock states given in Eq. (\ref{M32})
also explicitly satisfy the Schr\"{o}dinger equation \cite{nwh}.

The consequence in Eq. (\ref{q46}) is not surprising, but is due on account of the fact that the phases
in Eqs. (\ref{20}) and (\ref{21}) are defined via the use of the Schr\"{o}dinger equation.
Let us briefly check about this.
The substitution of the wave function $| \Psi_{\rm coh}(t) \rangle$ in Eq. (\ref{32w})
into the Schr\"{o}dinger equation gives
\be
| A \rangle \dot{\gamma}(t) = i \f{\p | A \rangle}{\p t} -\f{\hat{H}}{\hbar}| A \rangle .
\ee
By operating $\langle A|$ in each side in the above equation from left and taking an
integration with respect to $t$, we have
\ba
\gamma(t)-\gamma(t_0) &=& \int_{t_0}^t \langle A(t')
|i\frac{\partial}{\partial t'}| A(t') \rangle dt' \nonumber \\
& &- \f{1}{\hbar} \int_{t_0}^t \langle A(t') |
\hat{H}(\hat{q},\hat{p},t')| A(t') \rangle dt'.
\ea
We see that the first term in the right-hand side of the above equation is the geometric phase accumulated
during a time interval $t-t_0$,
whereas the second term is the dynamical phase evolved at the same time.
Although the total phase follows the Schr\"{o}dinger equation in this way, such compliance
requires the relation in Eq. (\ref{dft}) as a necessary condition: one can easily confirm that
the adopted time function $f(t)$, Eq. (\ref{4}), satisfies this condition exactly.

The theory of nonstatic waves along this line has been developed based on the fact that
the associated wave functions in Fock, coherent, and other quantum states
evolve in time as per the Schr\"{o}dinger equation.
Such evolutions in a static environment require the emergence or modification of the
geometric phase in the wave.
The phase $\gamma (t)$ is often neglected in the representation of the coherent wave
because it does mostly not affect the expectation value of an observable.
For instance, recent studies for the coherent states
in Refs. \cite{ncs,cox1,cox2} are the cases where $\gamma(t)$ is not considered.
However, $\gamma (t)$ and the geometric phase as its component play crucial
roles in some cases such as wave interferences and quantum superpositions.
The geometry-oriented phase difference
in the interference experiment is non-intuitive in many cases and
can only be understood via the geometric phase perspective.
Interpretation of shifts in the location of the wave maximum for
superposed waves, which are caused by geometric phase, demands
a rigorous analysis in general.
%%%%%%%%%%%%%%%%%%%%%%%%%%%%%%%%%%%%%%
\begin{figure}%[H]
\centering
\includegraphics[keepaspectratio=true]{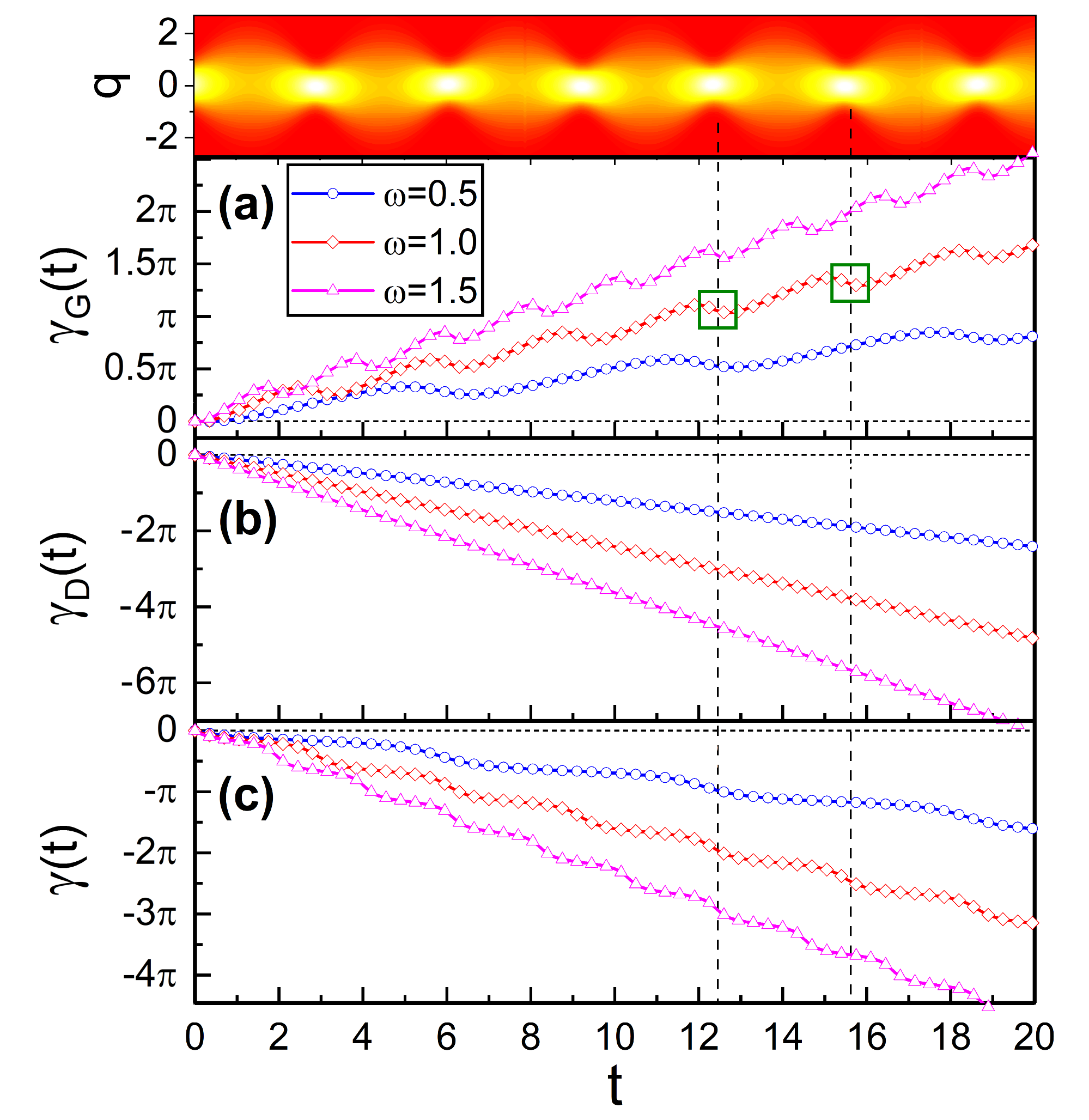}
\caption{\label{Fig3} Time evolution of the geometric phase ({\bf a}), dynamical phase ({\bf b}), and
total phase ({\bf c}) in the coherent state for several values of $\omega$.
For convenience of the related interpretation, density plot for the time
evolution of the corresponding probability density
with $\omega=1$ is shown in the uppermost part.
We have chosen the value of $A_0$ instead of $Q_0$, i.e., we setted $A_0=0.1$: we will also
specify the value of $A_0$ (not $Q_0$) in the subsequent figures for convenience from now on.
We have used $c_1=2.5$, $c_2=0.5$, $t_0=0$, $\varphi=\theta=0$, and $\gamma(t_0)=0$ (that is,
$\gamma_G(t_0)=\gamma_D(t_0)=0$).
 }
\end{figure}
%%%%%%%%%%%%%%%%%%%%%%%%%%%%%%%%%%%%%%
%%%%%%%%%%%%%%%%%%%%%%%%%%%%%%%%%%%%%%
\begin{figure}%[H]
\centering
\includegraphics[keepaspectratio=true]{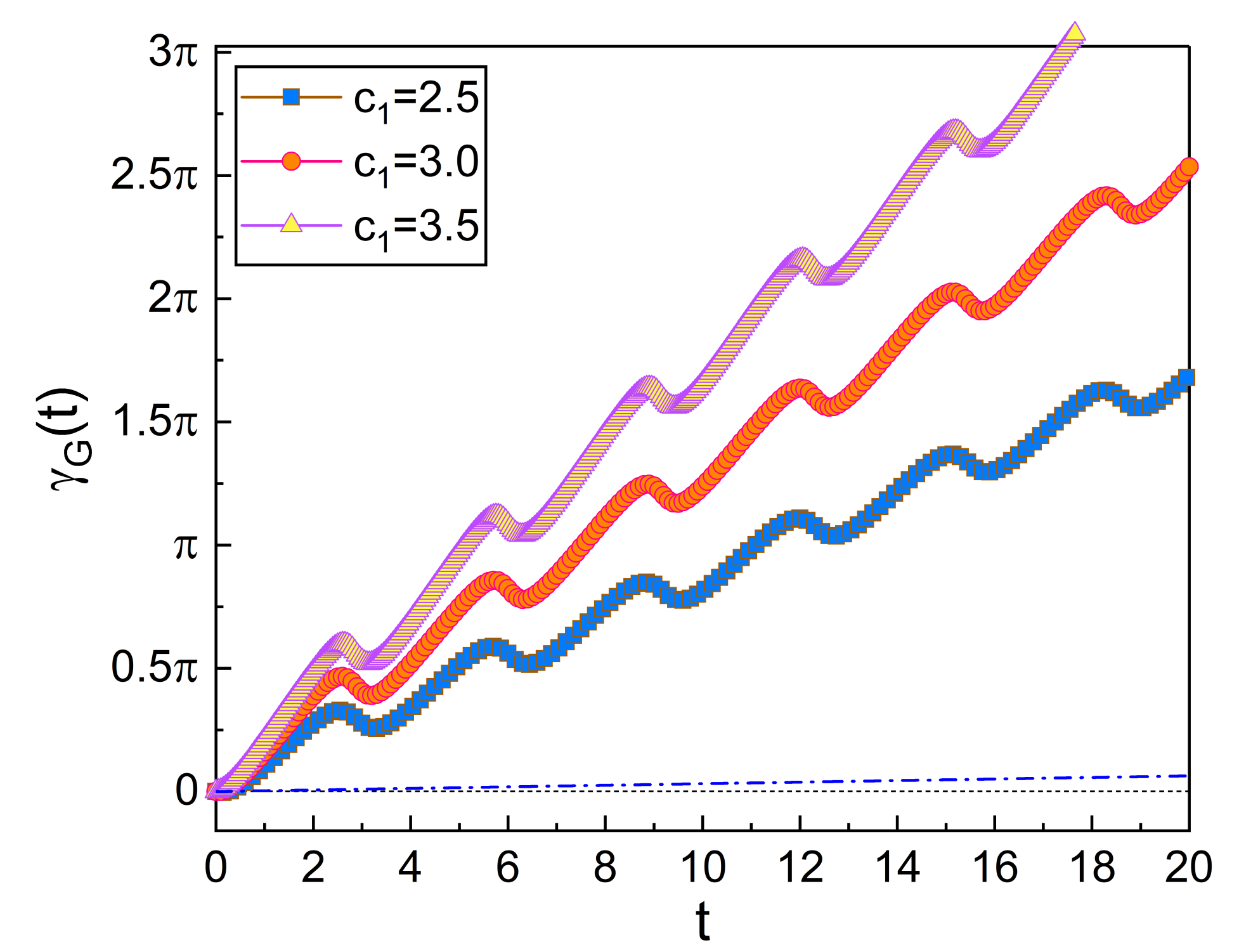}
\caption{\label{Fig4} Time evolution of the geometric phase in the coherent state for several
values of $c_1$.
Because wave nonstaticity is governed by $c_1$ (and $c_2$), we can confirm the
effects of nonstaticity on geometric phase from this figure.
We have used $\omega=1$, $A_0=0.1$, $c_2=0.5$ (except for an extra line),
$t_0=0$, $\varphi=\theta=0$, and $\gamma_G(t_0)=0$.
The measure of nonstaticity is 0.79, 1.02, and 1.22 in turn for the graphs
designated in the legend.
An extra line (dash-dot line) is the geometric phase without wave nonstaticity,
i.e., for the case where $c_1=c_2=1$ is chosen; this shows linear increase of the
phase as mathematically turned up from Eq. (\ref{27+1}).}
\end{figure}
%%%%%%%%%%%%%%%%%%%%%%%%%%%%%%%%%%%%%%
\\
\\
{\bf %5. Time Behavior of the Geometric Phase
5. ANALYSIS OF THE BEHAVIOR OF GEOMETRIC PHASE
\vspace{0.2cm}} \\
{\bf 5.1. Nonlinear Characteristics}
\\
We now clarify the influence of wave nonstaticity on phase evolutions
by analyzing the evolution of the geometric phase and its relation with the
dynamical phase together with their collaborative formation of
the resultant overall phase.
The time evolution of the geometric phase, the dynamical phase, and the overall
phase are plotted in Fig. 3 for several different values of $\omega$.
Roughly speaking, the geometric phase increases with time, while the dynamical phase decreases.
Because the absolute value of
the dynamical phase is greater than the geometric phase, the total phase decreases.
The increase of the geometric phase is more rapid when $\omega$ is large.

While the previously reported geometric phase \cite{snb,5} in the coherent state,
which is the same as Eq. (\ref{27+1}), increases monotonically over time, the geometric phase in this work
fluctuates as time goes by provided that $f(t) \neq 1$.
This is purely due to the effects of the wave nonstaticity on the evolution of the geometric phase.
From a careful analysis of panel ({\bf a}) in Fig. 3, we see that the geometric phase abruptly drops
whenever the wave forms a node in its evolution in the quadrature space (see the two rectangular regions
in Fig. 3({\bf a})).
Such a fluctuational variation of the phase also becomes rapid as $\omega$ increases.

%%%%%%%%%%%%%%%%%%%%%%%%%%%%%%%%%%%%%%
\begin{figure}%[H]
\centering
\includegraphics[keepaspectratio=true]{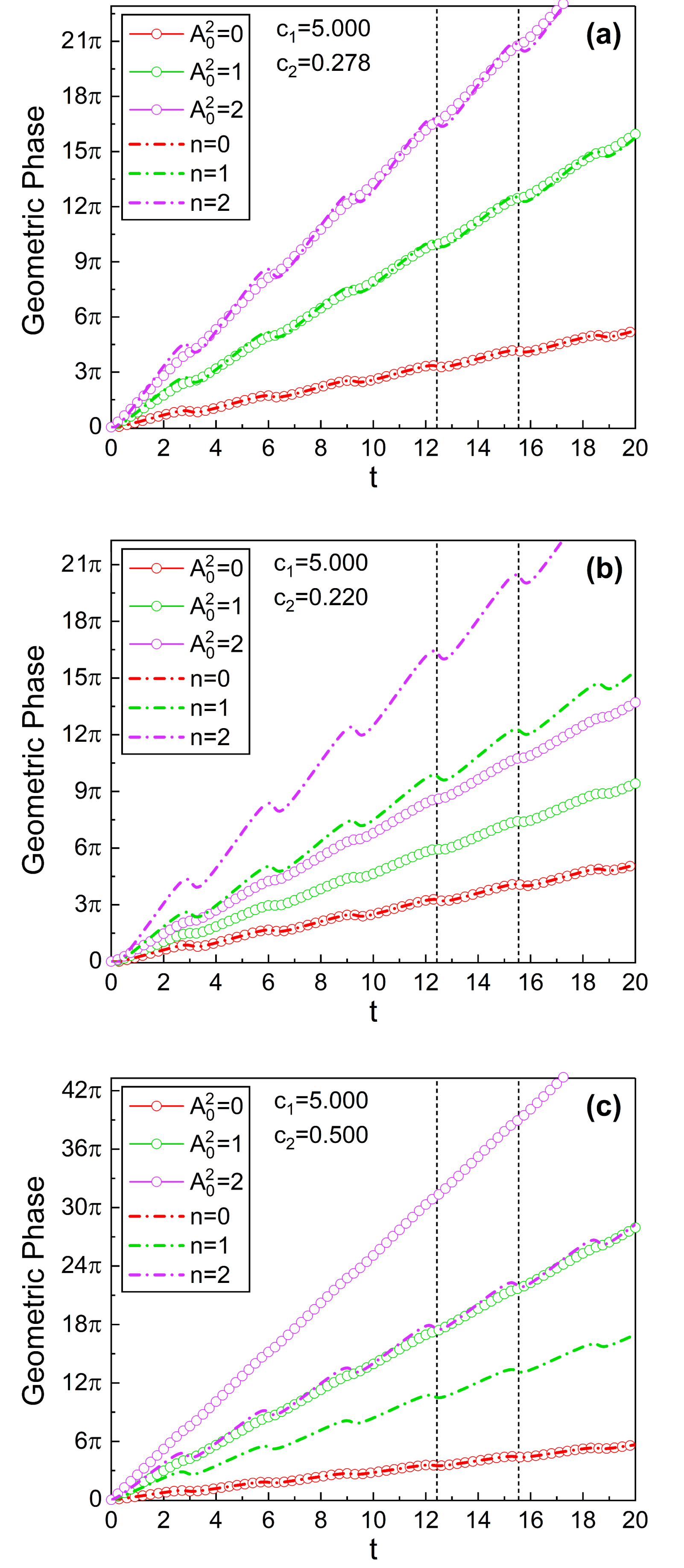}
\caption{\label{Fig5} Comparison of the evolution of the
geometric phase ($\gamma_G(t)$, circles) in the coherent state
with that ($\gamma_{G,n}(t)$, dash-dot lines) in the Fock states
for $A_0^2=n=0$ (red), $A_0^2=n=1$ (green), and $A_0^2=n=2$ (violet).
Two dotted vertical guidelines in each panel correspond to two adjacent nodes
in the evolution of the nonstatic coherent wave like those in Fig. 3:
this rule is also applied in subsequent figures up to Fig. 8.
The values of ($c_1$, $c_2$) are (5.000, 0.278) for ({\bf a}),
(5.000, 0.220) for ({\bf b}), and (5.000, 0.500) for ({\bf c}).
We have used $\omega=1$, $t_0=0$, $\varphi=\theta=0$,
and $\gamma_G(t_0)=\gamma_{G,n}(t_0)=0$.
The measure of nonstaticity is 1.73, 1.70, and 1.81 for the graphs in
panels ({\bf a}), ({\bf b}), and ({\bf c}), respectively.
}
\end{figure}
%%%%%%%%%%%%%%%%%%%%%%%%%%%%%%%%%%%%%%

We see the effects of $c_1$ on the evolution of the geometric phase from Fig. 4.
Because $c_1$ (together with $c_2$) is a factor that governs the measure of nonstaticity,
the increment of the geometric phase becomes large as $c_1$ grows.
Similar effects can also be confirmed when $c_2$ increases
instead of $c_1$ (or when both $c_1$ and $c_2$ increase).
Additionally, from the periodic pattern in the variation of three graphs in Fig. 4,
we confirm that wave nonstaticity is revealed via the nonlinearity
in the evolution of the geometric phase.
Such a variation naturally disappears in the static limit (see the dash-dot line in
lower part of Fig. 4).
%%%%%%%%%%%%%%%%%%%%%%%%%%%%%%%%%%%%%%
\begin{figure}%[H]
\centering
\includegraphics[keepaspectratio=true]{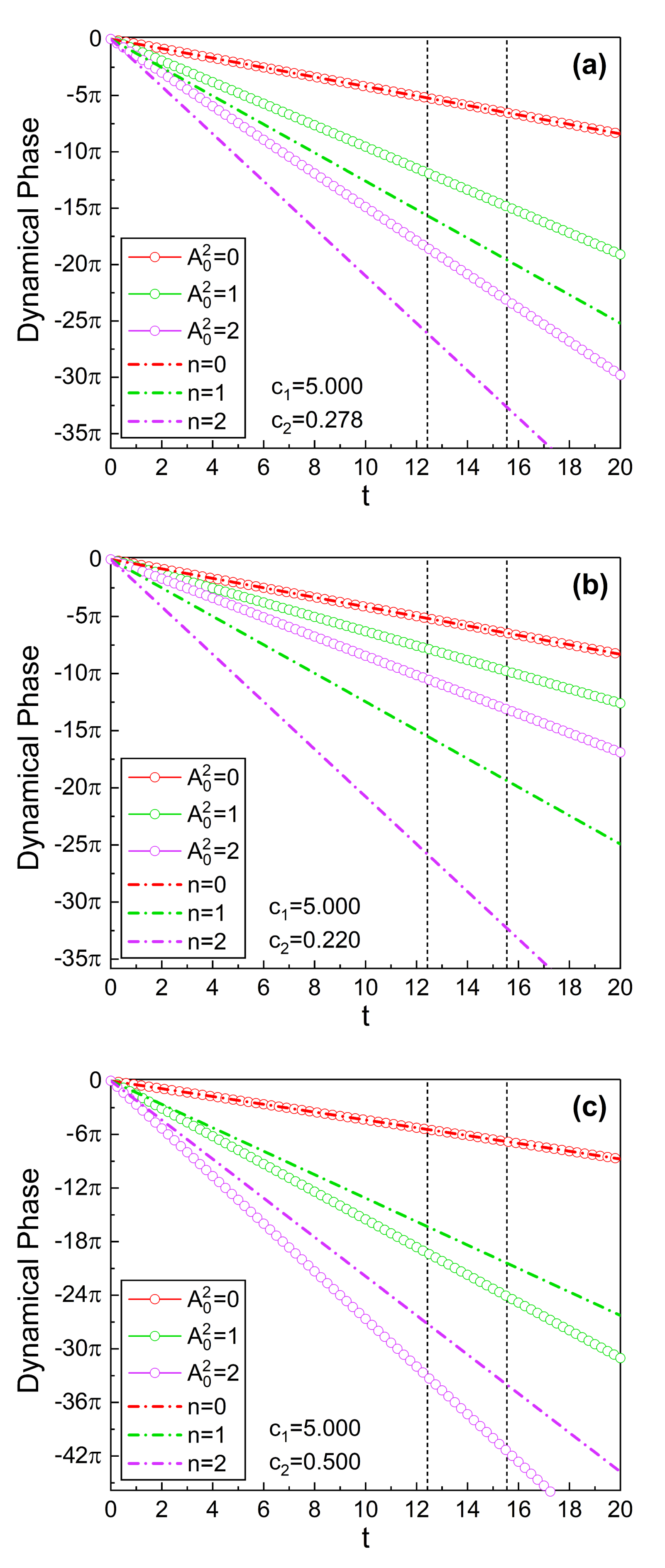}
\caption{\label{Fig6} Comparison of the evolution of the dynamical
phase ($\gamma_D(t)$, circles) in the coherent state
with that ($\gamma_{D,n}(t)$, dash-dot lines) in the Fock states
for $A_0^2=n=0$ (red), $A_0^2=n=1$ (green), and $A_0^2=n=2$ (violet).
The values of ($c_1$, $c_2$) are the same as those in Fig. 5 in turn
for ({\bf a}), ({\bf b}), and ({\bf c}), while
$\gamma_D(t_0)=\gamma_{D,n}(t_0)=0$ has been used.
All of the other chosen parameters are also the same as those in Fig. 5.}
\end{figure}
%%%%%%%%%%%%%%%%%%%%%%%%%%%%%%%%%%%%%%
\\
{\bf 5.2. Similarities and Differences with the Fock-State Phases}
\\
{\it 5.2.1. Harmonization with the dynamical phase}
\\
It may be worthwhile to compare the geometric phase in the coherent state to
that in the Fock states considering the similarity of Eq. (\ref{gDntr-}) with
Eq. (\ref{gDntr}), which shows that $A_0^2$ in the coherent state plays
a role similar to $n$ in the Fock states.
This consequence is closely related to the fact that
$\langle A | \hat{A}^\dagger \hat{A} | A \rangle = A_0^2$ whereas
$\langle \Phi_n | \hat{A}^\dagger \hat{A} | \Phi_n \rangle = n$.
Regarding this, the evolution of the geometric phase in the coherent state
is compared  to that in the Fock states in Fig. 5 for several different values of $c_2$
while $c_1$ is taken to be a fixed value which is 5.000.
This figure shows that the geometric phase in the coherent state increases as
$A_0^2$ grows, whereas that in the Fock states increases as $n$ grows.
The average rate of such an increase in the coherent state is large when $c_2$
is large, while that in the Fock states is independent of $c_2$
in the case adopted in Fig. 5: this fact can be confirmed by
comparing panels ({\bf a}), ({\bf b}), and ({\bf c}) in Fig. 5 with one another
regarding the differences in the scale of vertical axes.
Figure 5({\bf a}) exhibits that the average rate of the increment of $\gamma_G(t)$ is the same as that
of $\gamma_{G,n}(t)$ when $c_2$ is a specific value, which is 0.278.
Figure 5({\bf b}) (Fig. 5({\bf c})) shows that the average rate of the increment
of $\gamma_G(t)|_{A_0^2=i}$ ($i=1,2$) is smaller (larger) than that
of $\gamma_{G,n}(t)|_{n=i}$ provided that $c_2$ is smaller (larger) than $0.278$.

On the other hand,
we see from three panels of Fig. 5 that, for the case of $A_0^2=n=0$, $\gamma_G(t)$ is
the same as $\gamma_{G,n}(t)$.
It is also possible to show this fact analytically under the choice $\varphi =\theta =0$.
If we insert the formula of $\Gamma_D (\bar{t})$ given in Eq. (\ref{vt0}) into
Eq. (\ref{gG}) with the condition $A_0=0$, the geometric phase becomes
\be
\gamma_G(t) = - \f{1}{2} \omega T(t) + \omega \f{c_1 + c_2}{4} [t-t_0]+ \gamma_G(t_0).
\label{ndcg}
\ee
This is the same as the zero-point Fock-state geometric phase
$\gamma_{G,0}(t)$ that can be obtained from Eq. (\ref{gGnt}) by setting
$n=0$ in addition to imposing a trivial condition $\gamma_G(t_0) = \gamma_{G,0}(t_0)$.
For the case of $c_1 = c_2 \rightarrow 1$, the first term in the right hand
side of Eq. (\ref{ndcg}) becomes $- \omega (t-t_0)/2$,
leading to $\gamma_G(t)$ being zero provided that $\gamma_G(t_0)=0$ as
the initial condition.
This outcome implies that the geometric phase disappears when its two sources are removed,
i.e., it disappears when both the displacement and the nonstaticity are removed.

%%%%%%%%%%%%%%%%%%%%%%%%%%%%%%%%%%%%%%
\begin{figure}%[H]
\centering
\includegraphics[keepaspectratio=true]{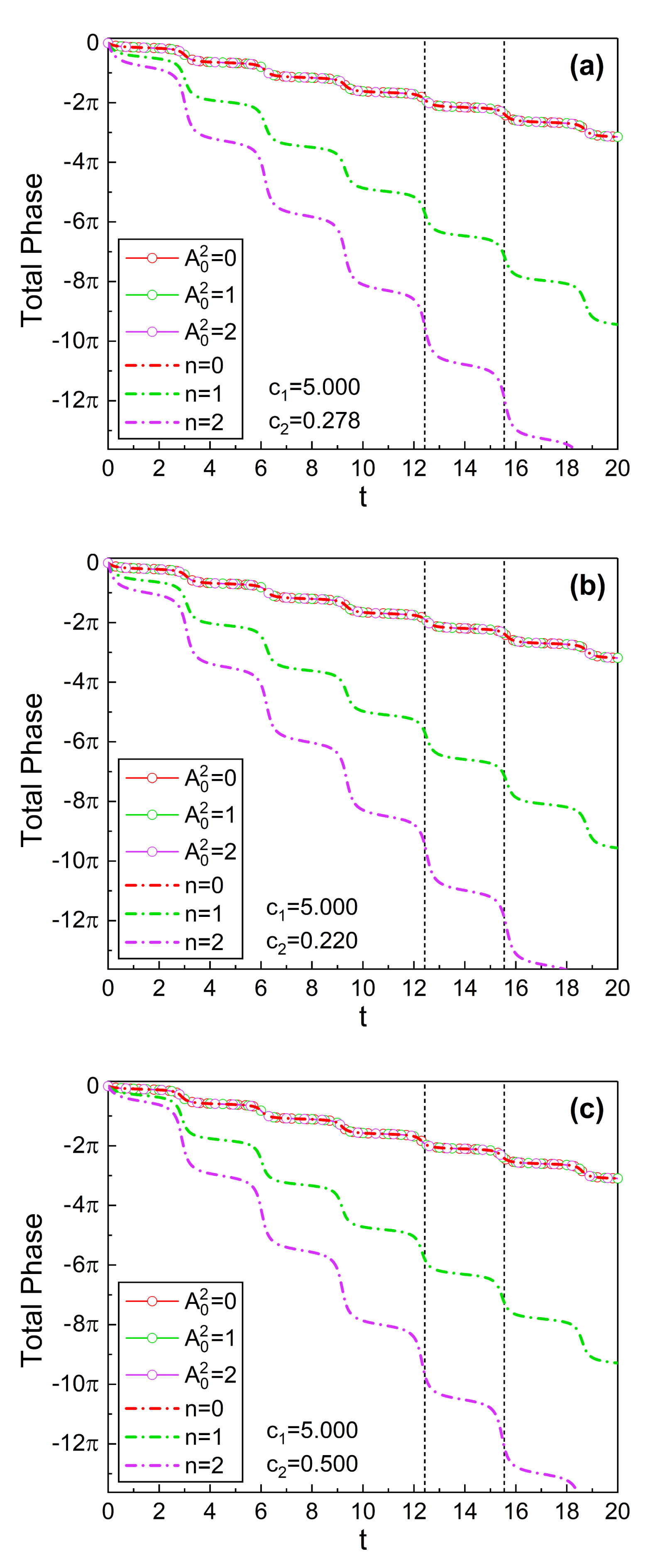}
\caption{\label{Fig7} Comparison of the evolution of
the total phase ($\gamma(t)$, circles) in the coherent state
with that ($\gamma_n(t)$, dash-dot lines) in the Fock states
for $A_0^2=n=0$ (red), $A_0^2=n=1$ (green), and $A_0^2=n=2$ (violet).
The values of ($c_1$, $c_2$) are the same as those in Fig. 5 in
turn for ({\bf a}), ({\bf b}), and ({\bf c}), while
$\gamma(t_0)= \gamma_n(t_0)=0$ has been used.
All of the other chosen parameters are also the same as those in Fig. 5.}
\end{figure}
%%%%%%%%%%%%%%%%%%%%%%%%%%%%%%%%%%%%%%

Figure 6 is the comparison of the evolution of $\gamma_D(t)$ with that
of $\gamma_{D,n}(t)$, for several different values of $c_2$ like in
the case of Fig. 5.
$\gamma_D(t)$ is always the same as $\gamma_{D,n}(t)$ for the case of $A_0^2=n=0$:
this consequence is quite similar to the case of the geometric phases
and its analytical verification can also be done by the same way
adopted in the case of those phases.
However, the absolute values of $\gamma_D(t)|_{A_0^2=i}$ ($i=1,2$) are smaller than
those of $\gamma_{G,n}(t)|_{n=i}$ for the case of
Figs. 6({\bf a}) ($c_2=0.278$) and 6({\bf b}) ($c_2=0.220$),
while the absolute values of $\gamma_D(t)|_{A_0^2=i}$ are larger than
those of $\gamma_{G,n}(t)|_{n=i}$ for the case of Fig. 6({\bf c}) ($c_2=0.500$).

%%%%%%%%%%%%%%%%%%%%%%%%%%%%%%%%%%%%%%
\begin{figure}%[H]
\centering
\includegraphics[keepaspectratio=true]{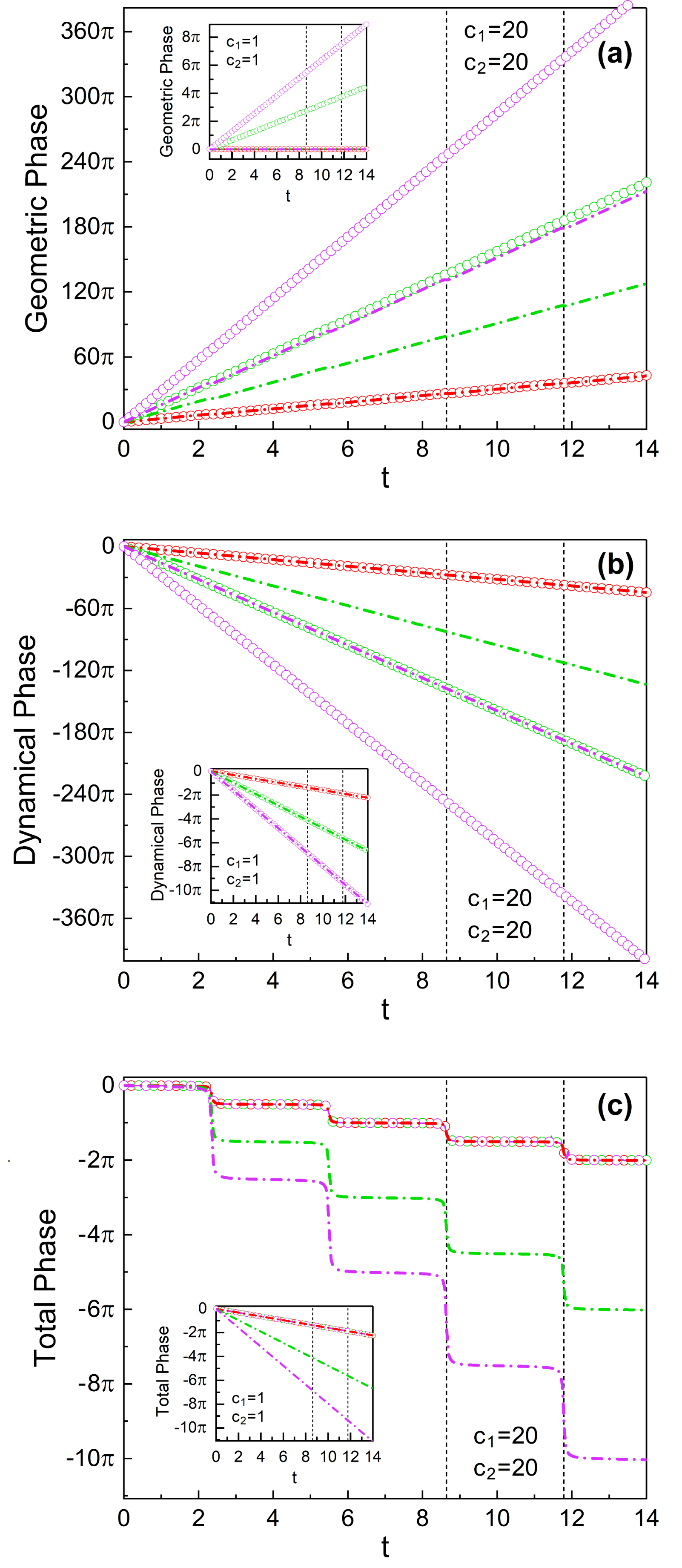}
\caption{\label{Fig8} The evolution of phases
in the coherent (circles) and Fock (dash-dot lines) states
for a light that has extreme nonstaticity: we have chosen
($c_1$, $c_2$)$=$($20$, $20$) and this choice corresponds to the
nonstaticity measure of $D= 14.12$.
({\bf a}) is for $\gamma_{G}(t)$ and $\gamma_{G,n}(t)$,
({\bf b}) is for $\gamma_{D}(t)$ and $\gamma_{D,n}(t)$,
and ({\bf c}) is for $\gamma(t)$ and $\gamma_n(t)$.
The inset in each panel is in the limit of the static light-waves with the choice
($c_1$, $c_2$)$=$($1$, $1$) for comparison.
The rule for the values of $A_0^2$ and $n$ is the same as that in Figs. 5, 6, and 7:
that is, we choose $A_0^2=n=0$ (red), $A_0^2=n=1$ (green), and $A_0^2=n=2$ (violet).
All chosen values of other parameters are also the same as those in Figs. 5, 6, and 7.}
\end{figure}
%%%%%%%%%%%%%%%%%%%%%%%%%%%%%%%%%%%%%%

Figure 7 is the evolutions of the total phases, where the values of the nonstaticity measure
are the same as those of Fig. 5, which are not so different from panel to panel.
This figure shows that the dependency of the total phase on $c_2$ is negligible if
the measures of nostaticity are not much deviate from one another.
However, since the phases in the nonstatic-wave evolutions are sensitive to the values of
$c_1$ and $c_2$ (for example, compare the phases of light waves in
the three panels of Fig. 2 in
Ref. \cite{nwh}), the phase $\gamma(t)$ (and $\gamma_n(t)$) in Fig. 7 is
a little different from panel to panel.
The phase in Fig. 7({\bf b}) (Fig. 7({\bf c})) is slightly
moved to the left (right) compared with that of Fig. 7({\bf a}).
Of course, these minute phase shifts are also found in the geometric phase
represented in Fig. 5.
However, the scale of $c_1$ and $c_2$ does not affect the period of
the nonlinear term, which is the first term in the right hand side of
Eqs. (\ref{gnl}) and (\ref{gnl-}).
Such a period is dependent only on $\omega$ as can be seen from Fig. 3.
That is, the period of the nonlinear evolution is small when $\omega$ is large
and large when $\omega$ is small.

The similarity in the evolutions of the envelopes of the total phases
in the three panels of Fig. 7 is noticeable,
if we consider that the corresponding geometric-phase evolutions shown in Fig. 5
are very different from panel to panel.
This consequence means that the evolutions of the geometric phases are
harmonized with those of the dynamical phases in a way that the total phases are
adjusted appropriately.
As a consequence, the total phase in the coherent state
is irrelevant to $A_0$ and always the same as that in the zero-point
Fock state ($n=0$).
\\
{\it 5.2.2. Extreme nonstaticity limits}
\\
It may be instructive to see the phase evolutions for higher nonstatic waves,
in addition to those that we have already seen from Figs. 5, 6, and 7.
The phase evolutions with an
extreme nonstaticity are represented in Fig. 8 with the choice of $c_1=c_2=20$.
While the measure of nonstaticity in this case is $D=14.12$,
the evolutions of phases in the static-wave limit ($D=0$) are also given in
insets in Fig. 8 for a comparison purpose.
Figure 8({\bf a}) exhibits that the geometric phase evolves nearly linearly under this
gigantic nonstaticity.
Although the nonlinear term $\gamma_{G, {\rm NL}}(t)$ in Eq. (\ref{gnl}) is being
seriously distorted as the measure of nonstaticity increases,
its contribution to the total phase becomes relatively small with the grow
of the nonstaticity because
the scale of the linear term $\gamma_{G, {\rm L}}(t)$ highly increases at the same tine.
$\gamma_{G, {\rm L}}(t)$ becomes dominant in this way
and, thereby, the nonlinear-term-induced distortion
in the geometric-phase evolution is buried in the
linear-term outcome as the nonstaticity becomes such a very high value.
This is the reason why the graphs in Fig. 8({\bf a}) looks like almost linear
in spite of the significant distortion of the part associated to its nonlinear-term.
The geometric phase in the coherent state is not
zero even when the nonstaticity disappears, except for the case of $A_0^2=0$,
as we have confirmed previously.
However, the inset in Fig. 8({\bf a}) exhibits that the geometric
phases in the Fock states are in contrast always zero
provided that the nonstaticity disappears.
Recall that the wave nonstaticity is the only source of appearing the geometric phase
for the case of the non-displaced Fock states considered here.

Figure 8({\bf b}) shows that the increase of the absolute values of the dynamical phases over time
is conspicuously small for the case of Fock states compared to that of the coherent state.
However, we confirm from the inset in Fig. 8({\bf b}) that that increment becomes
the same as that in the
coherent state in the static-wave limit: notice that the structural equality
between Eqs. (\ref{gDntr-}) and (\ref{gDntr}) mathematically supports this consequence.
It looks like, from Fig. 8({\bf b}), that $\gamma_D(t)$ with $A_0^2=1$ is
the same as $\gamma_{D,n}(t)$ with $n=2$.
To see this outcome in detail, we consider the approximate value of the geometric phase
in the coherent state utilizing Eq. (\ref{gD}) with Eq. (\ref{vt0})
in the limit $c_1 = c_2 \gg 1$ with $\gamma_D(t_0)=0$:
\be
\gamma_D(t) \approx - \omega \bigg(2A_0^2 +\f{1}{2}\bigg) c_1 (t-t_0) ,
\ee
which is the same as the exact value of $\gamma_{D,n}(t)$ given in Eq. (\ref{gDnt})
in the same limit, provided that $A_0^2 = n/2$ and $\gamma_{D,n}(t_0)=0$.
Hence, $\gamma_D(t)$ with the choice $A_0^2=1$ in the case of a highly nonstatic wave
is very the same as $\gamma_{D,n}(t)$ with $n=2$ as also confirmed graphically.

%%%%%%%%%%%%%%%%%%%%%%%%%%%%%%%%%%%%%%
\begin{figure}%[H]
\centering
\includegraphics[keepaspectratio=true]{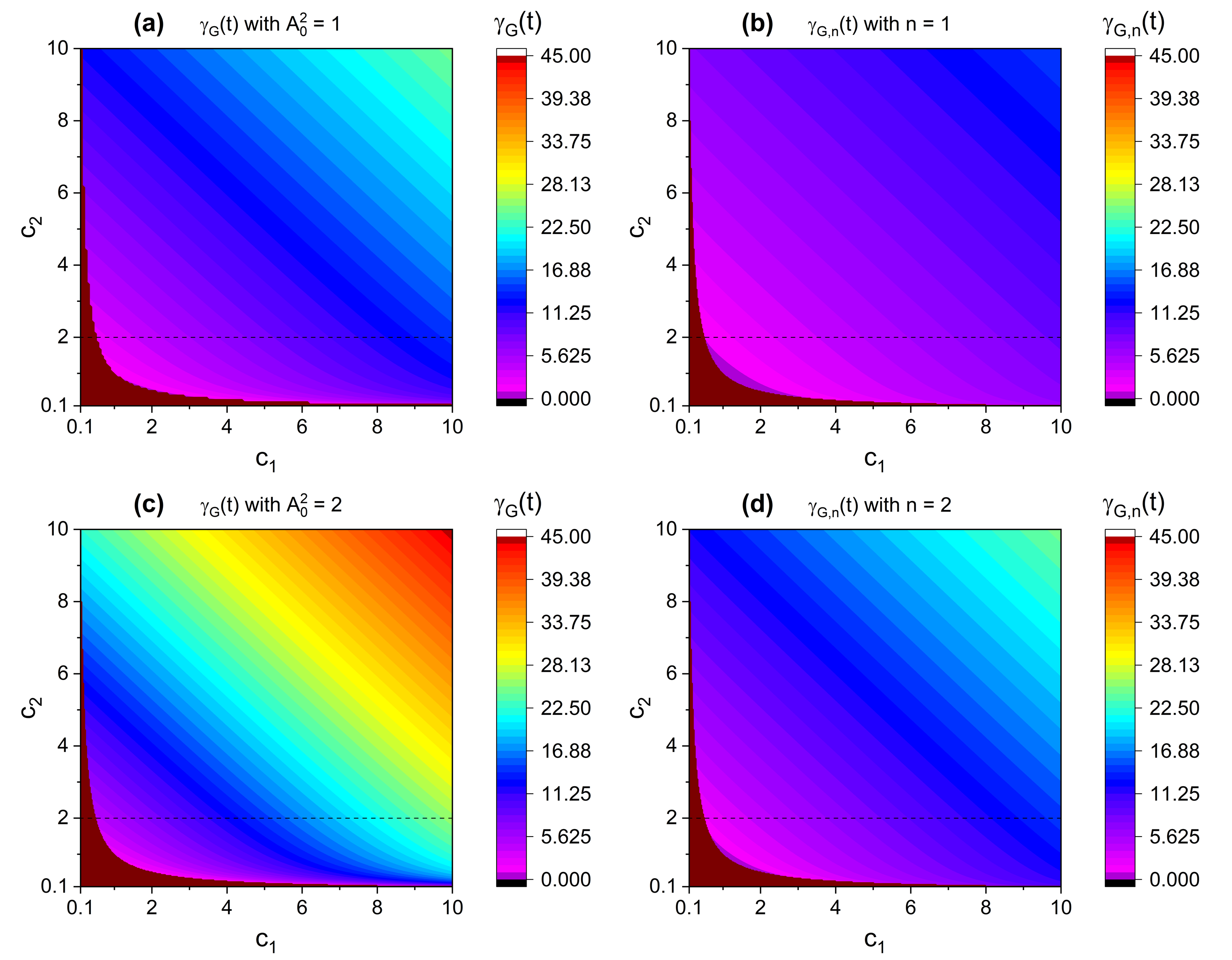}
\caption{\label{Fig9} Density plots for the geometric phases in the coherent and Fock states
depending on $c_1$ and $c_2$:
({\bf a}) is $\gamma_G(t)$ with $A_0^2 =1$, ({\bf b}) is $\gamma_{G,n}(t)$ with $n =1$,
({\bf c}) is $\gamma_G(t)$ with $A_0^2 =2$, and ({\bf d}) is $\gamma_{G,n}(t)$ with $n =2$.
We considered the geometric phases evolved during a unit time by choosing $t_0=0$ and $t=1$.
The dark reddish brown region in the lower left corner of each panel is
invalid/noncalculable region according to the condition given in Eq. (\ref{c5+1}).
We used $\omega=1$, $\varphi=\theta=0$, and $\gamma_G(t_0)=0$.
}
\end{figure}
%%%%%%%%%%%%%%%%%%%%%%%%%%%%%%%%%%%%%%

%%%%%%%%%%%%%%%%%%%%%%%%%%%%%%%%%%%%%%
\begin{figure}%[H]
\centering
\includegraphics[keepaspectratio=true]{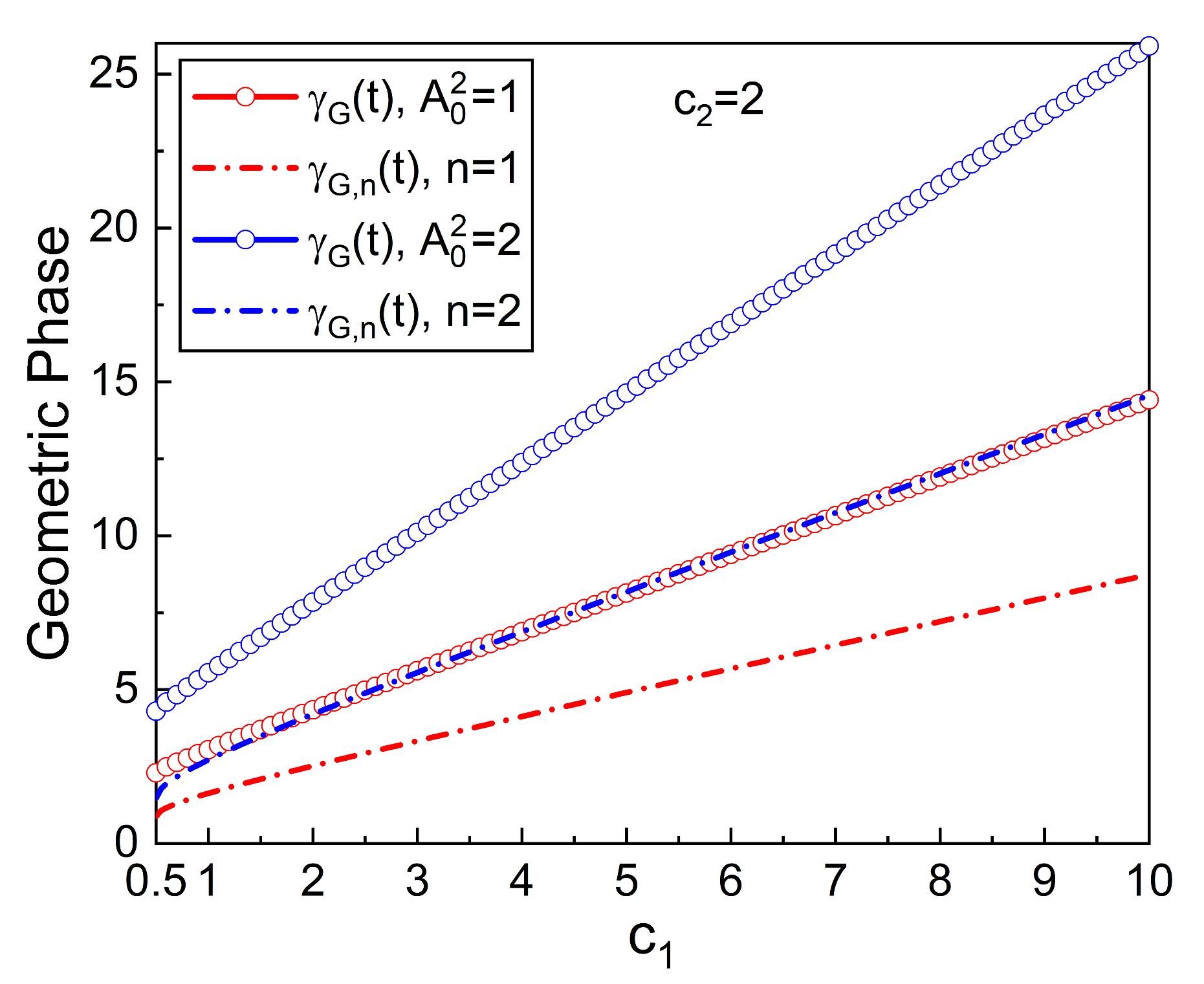}
\caption{\label{Fig10} Comparison of the geometric phase along the dashed
lines in each panel of Fig. 9.
That is, the variation of the geometric phases depending on $c_1$
for four specific cases where we have chosen $c_2=2$:
they are $\gamma_{G}(t)$ with $A_0^2=1$ (red circle line) and $A_0^2=2$ (blue circle line),
and $\gamma_{G,n}(t)$ with $n=1$ (red dash-dot line) and $n=2$ (blue dash-dot line).
We considered only the valid value as the
initial value of $c_1$ in the horizontal axis
according to the restriction of Eq. (\ref{c5+1}): $c_1 \geq 0.5$.
We used the values of parameters given in Fig. 9, i.e.,
$\omega=1$, $t_0=0$, $t=1$, $\varphi=\theta=0$, and $\gamma_G(t_0)=0$.
}
\end{figure}
%%%%%%%%%%%%%%%%%%%%%%%%%%%%%%%%%%%%%%

The harmonized evolution of the dynamical phases given in Fig. 8({\bf b})
makes the total phases (Fig. 8({\bf c})) similar to those in Fig. 7,
except for the prominence of their strange nonlinear behavior.
We see from Fig. 8({\bf c}) that the total phases abruptly drop at the nodes in
the periodic nonstatic-wave evolutions.
Whereas their evolving patterns are fairly
different compared to those in Fig. 7, this outcome is evidently due to the
greatness of the measure of nonstaticity.
Such precipitations in the graphs however entirely disappear in the static-wave limit,
forming their evolutions being linear (see inset of Fig. 8({\bf c})).
By the way, the average gradient in the evolution of the total phase and
the period of its nonlinear portion are different depending on $\omega$
as it can be checked from Fig. 3({\bf c}).
%%%%%%%%%%%%%%%%%%%%%%%%%%%%%%%%%%%%%%
\begin{figure}%[H]
\centering
\includegraphics[keepaspectratio=true]{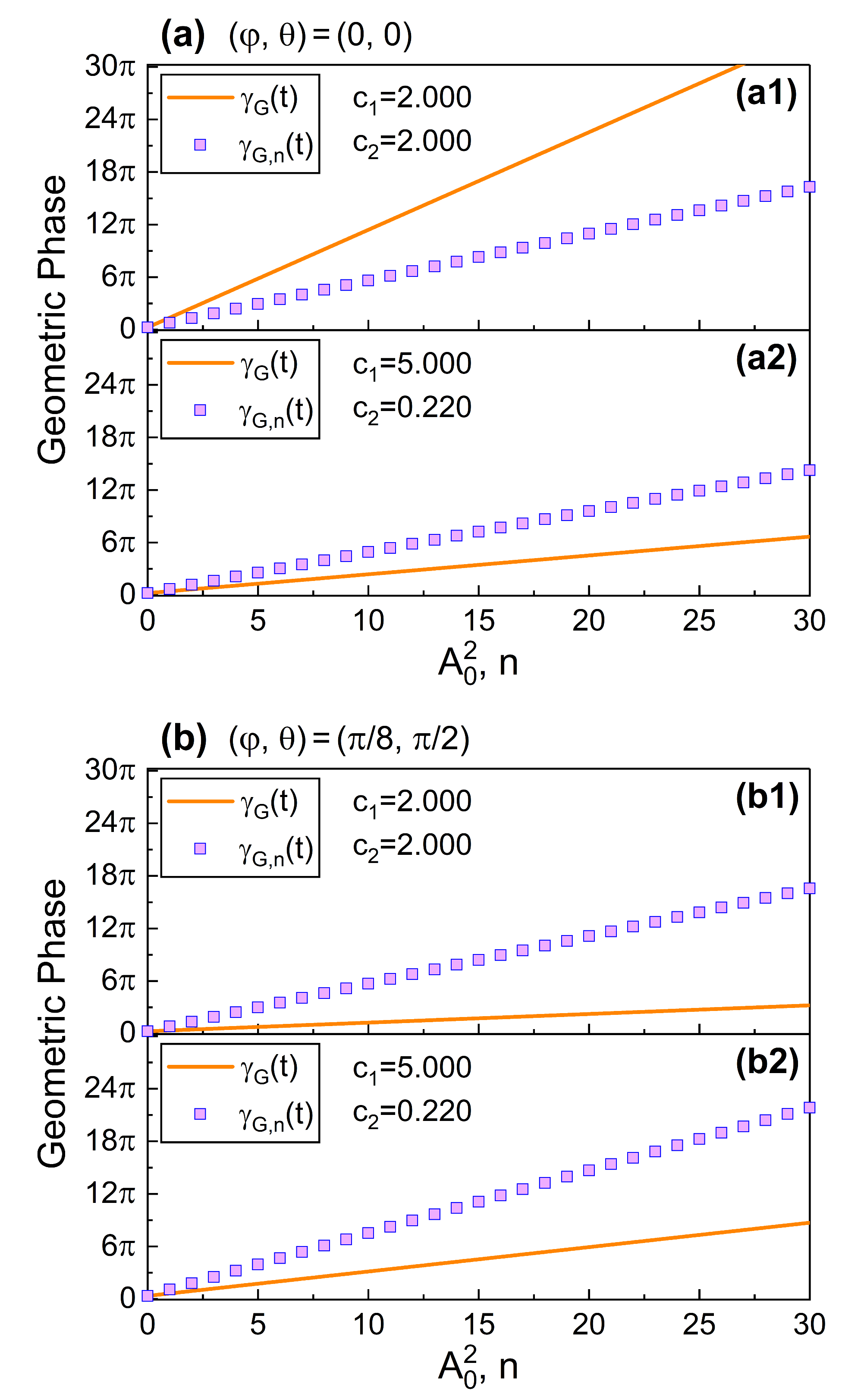}
\caption{\label{Fig10} The dependence of the geometric phases
$\gamma_{G}(t)$ and $\gamma_{G,n}(t)$ evolved over unit time
on $A_0^2$ (for $\gamma_{G}(t)$) and $n$ (for $\gamma_{G,n}(t)$),
where $(\varphi,\theta)=(0,0)$ for ({\bf a})
and $(\varphi,\theta)=(\pi/8,\pi/2)$ for ({\bf b}).
$t_0=0$ and $t=1$ have been taken in this figure to adopt the unit-time interval.
We have chosen $(c_1,c_2)=(2.000,2.000)$ for subpanels $(\bf a1)$ and $(\bf b1)$
and $(c_1,c_2)=(5.000,0.220)$ for $(\bf a2)$ and $(\bf b2)$,
while $\omega=1$ and $\gamma_G(t_0)=\gamma_{G,n}(t_0)=0$ are used.
%Other values used are .
}
\end{figure}
%%%%%%%%%%%%%%%%%%%%%%%%%%%%%%%%%%%%%%
\\
{\it 5.2.3. More detailed comparison with the Fock-state geometric phases}
\\
More detailed dependence of the geometric phase on $c_1$ and $c_2$
can be seen from Fig. 9.
We can see from this figure that the increase of the geometric phase
in a unit time is large when both $c_1$ and $c_2$ are high
for both the coherent and the Fock states.
Figure 10 is the comparison of the geometric phase along the dashed lines in Fig. 9.
All four graphs of the geometric phase in this figure
increase almost monotonically as $c_1$ grows.
We have also confirmed similar increase of the geometric phase along with the increase of $c_2$
(not shown).
This consequence means that the wave nonstaticity, which is determined by
$c_1$ and $c_2$, is the main factor that is
responsible for the advent of the geometric phase (or enhancement of an
existing geometric phase when, for example, $A_0 \neq 0$ in the coherent state).

By the way, Fig. 9 together with Fig. 10 shows that, as the degree of nonstaticity increases,
the geometric phase in the case of the coherent state is more dominant
rather than that in the Fock states: while this is the usual consequence
confirmable based on Fig. 9 for most choices of $c_1$ and $c_2$, an exception is the case
given in Fig. 5({\bf b}).
Regarding this consequence, we have shown the dependence of the geometric phase on
$A_0^2$ (for the coherent state) and $n$ (for Fock states) in Fig. 11.
Though $A_0^2$ in the coherent state plays the role similar to $n$ in the Fock states
in a sense that the increase of the geometric phase over time is
controlled by them almost in the same manner, the two factors
are physically different from each other as it is known.
$A_0^2$ determines the scale of displacement in the coherent state,
while $n$ determines energy level in the Fock states.
We conclude from Fig. 11({\bf a1}) that the geometric phase is usually
more significantly affected by the displacement rather than
the increase of the energy level when $\varphi=\theta=0$,
whereas Fig. 11({\bf a2}) shows an exceptional case.
Figure 11({\bf b}) however manifests that this consequence does not hold
if we choose other values of $\varphi$ and $\theta$.
Meanwhile, the geometric phase in a non-displaced coherent state ($A_0^2=0$)
is exactly the same as that in the zero-point Fock state provided that
$\varphi=\theta=0$ as previously expected according to Eq. (\ref{ndcg}).
\\
\\
%\section{Conclusion}
{\bf %6. Conclusion
6. CONCLUSION
\vspace{0.2cm}} \\
The geometric phase of a nonstatic light in the coherent state, arisen in a static environment,
has been investigated by adopting a generalized annihilation operator $\hat{A}$.
$\hat{A}$ is represented in terms of the nonlinear time function $f(t)$
that is controlled by the nonstaticity parameters $c_1$ and $c_2$.
We introduced a time-varying nonstatic wave function, which is the eigenfunction of $\hat{A}$.
Then, the geometric phase associated with the wave nonstaticity
was evaluated by utilizing that wave function.

Our general geometric phase
showed nonlinear variation due to the consideration of wave nonstaticity.
A regular pattern in the variation of the geometric phase has been recognized:
we have analyzed it in detail along with examining
the effects of the overall phase caused by the peculiar behavior of the geometric phase
as its component.
The evolution of the geometric phase in time is a combination of its linear increase and
a sinusoidal-like fluctuation.
The fluctuation in the evolving of phase is raised from the
nonlinearity of $f(t)$ accompanying the wave nonstaticity.
The increment of the geometric phase over time becomes rapid when
$c_1$ and $c_2$ become large.
Such increment in the coherent state is quite drastic in most cases compared to
that in the Fock states.
We confirmed that the angular frequency $\omega$ and amplitude
$A_0$ also affect the pattern in the phase evolutions.
For instance, the frequency of the nonlinear portion in the
geometric-phase fluctuation is relatively high when $\omega$ is large.

The evolution of the geometric phase is harmonized with the dynamical phase,
leading to the time-averaged incrementing rate of the total phase being
not affected by the magnitude of wave nonstaticity.
Notably, the scale of distortion in the total phase is
determined only by the nonstaticity measure, while the average gradient of the
total phase and the period of its nonlinear portion are determined by $\omega$.
For the case of an extreme nonstaticity, the nonlinear portion of the geometric phase
undergoes a great distortion and, thereby, the total phase exhibits a unique and singular
pattern in its evolution.
That is, total phase abruptly drops whenever the wave forms a node periodically
over time.
However, in the limit of $f(t) =1$, the fluctuation of the geometric phase disappears,
showing only its linear increase.
The remaining rate of the geometric-phase increase over time corresponds only to the
elemental part in this limit.
Namely, the nonstaticity-induced portion in such a linear increment no longer exists.

It was demonstrated that the nonstatic wave function in the coherent state satisfies
the Schr\"{o}dinger equation
only when we consider the geometric phase as an additional phase to the dynamical one.
This fact means that the emergence of the geometric phase is
essential when the wave exhibits nonstaticity.
The geometric phase is one of the most fundamental elements in quantum mechanics
in connection with a nonstatic-wave evolution and parallel transport of the state.
Exact mathematical formulation of the geometric phase in the generalized
coherent state achieved in this work is necessary
not only for extending our knowledge in this field, but for pursuing further
studies in relation with nonstatic light rigorously as well.
Characterizing profound properties of the geometric phase is
important for its substantial utilization in diverse emerging scientific fields,
such as the quantum information technology, metamaterials-based
photonic devices, light control for Holograms,
and detectors for gravitational waves \cite{nco1,nco2,nco3}.

\appendix
\section{Full version of the coherent-state expansion over Fock states}
It is possible to represent the wave function in the coherent state in the manner of Eq. (\ref{qA-1}) instead of  Eq. (\ref{qA}).
That is, regarding Eqs. (\ref{M32}), (\ref{qA}), and (\ref{32w}), we have
\be
\langle q |\Psi_{\rm coh}(t) \rangle= \sum_{n=0}^\infty a_n \langle q |\Psi_n(t) \rangle ,  \label{AqA}
\ee
where $a_n$ in this case take the form
\ba
a_n &=& b_n (A) \exp \{- i [\gamma_n(t)-\gamma(t)] \} \nonumber \\
&=& \exp (-A_0^2/2) \f{A_0^n}{\sqrt{n!}} \exp \{-i[\gamma_n(t_0)-\gamma(t_0)]-in\theta \},
\label{AqA+1}
\ea
whereas $A_0$ and $\theta$ are constants over time, which appear in Eq. (\ref{18}).
In the derivation of the last formula in Eq. (\ref{AqA+1}), we used Eqs. (\ref{18}),
(\ref{32}), and (\ref{gnt}).
Note that Eq. (\ref{AqA+1}) does not vary in time.

\section{Gauge invariance of the geometric phase}
To show the gauge invariance property of the geometric phase,
let us consider a model of U1
gauge transformation for the full wave function $| \Psi_{\rm coh}(t) \rangle$ given
in Eq. (\ref{32w}). We represent such a transformation as
\be
| \tilde{\Psi}_{\rm coh}(t) \rangle = e^{i\alpha(t)} | \Psi_{\rm coh}(t) \rangle,  \label{me35}
\ee
where $\alpha(t)$ is an arbitrary time-dependent phase.
To tackle our problem of the demonstration in connection
with Eq. (\ref{me35}), we rewrite the geometric phase
represented in Eq. (\ref{20}) such that
\cite{nmu}
\be
\gamma_{G}(t) =  {\rm arg} (\langle \Psi_{\rm coh}(t_0)| \Psi_{\rm coh}(t) \rangle)+\int_{t_0}^t \langle \Psi_{\rm coh}(t')
|i\frac{\partial}{\partial t'}| \Psi_{\rm coh}(t') \rangle dt' +\gamma_{G}(t_0).
\ee
It is not difficult to prove that this formula of the geometric
phase is the same as Eq. (\ref{20}) using the relation given in Eq. (\ref{32w}).

Then the geometric phase related to $| \tilde{\Psi}_{\rm coh}(t) \rangle$
can be written in the form
\ba
\tilde{\gamma}_{G}(t) &=&  {\rm arg} (\langle \tilde{\Psi}_{\rm coh}(t_0)| \tilde{\Psi}_{\rm coh}(t) \rangle)
+\int_{t_0}^t \langle \tilde{\Psi}_{\rm coh}(t')
|i\frac{\partial}{\partial t'}| \tilde{\Psi}_{\rm coh}(t') \rangle dt' +\gamma_{G}(t_0) \nonumber \\
&\equiv& I_1 + I_2 +\gamma_{G}(t_0),
\ea
where $I_1$ and $I_2$ indicate merely the first and second terms
of the mathematical expression in the former line, respectively.
We have assumed $\tilde{\gamma}_{G}(t_0)=\gamma_{G}(t_0)$ in the above expression for simplicity.
Minor evaluations using Eq. (\ref{me35}) give
\ba
I_1 &=& {\rm arg} (\langle \Psi_{\rm coh}(t_0)| \Psi_{\rm coh}(t) \rangle) + \alpha(t) - \alpha(t_0), \\
I_2 &=& -\int_{t_0}^t \dot{\alpha}(t') dt' + \int_{t_0}^t \langle \Psi_{\rm coh}(t')
|i\frac{\partial}{\partial t'}| \Psi_{\rm coh}(t') \rangle dt'.
\ea
Using the relation $\int_{t_0}^t \dot{\alpha}(t') dt' = \alpha(t) - \alpha(t_0)$
in connection with $I_2$,
we now immediately confirm that $\tilde{\gamma}_{G}(t) = \gamma_{G}(t)$.
Thus the geometric phase is invariant under the considered
transformation of the gauge.

\section{Intermediate steps in evaluating phases }
We first represent
the methods for obtaining the coefficients $g_1(t)$-$g_4(t)$ appeared in Eqs. (\ref{24})-(\ref{27})
as the factors of the geometric phase.
By evaluating Eq. (\ref{20}) with the use of Eq. (\ref{22}),
the geometric phase becomes Eq. (\ref{23}) with
\ba
g_1(t)&=& -\f{1}{ f(t)}({A}^2 + {A}^{*2}-2A^*A+1), \label{34}  \\
g_2(t)&=&\f{[\dot{f}(t)]^2}{f(t)}({A}^2 + {A}^{*2}+2A^*A+1), \label{35}  \\
g_3(t)&=&\f{i\dot{f}(t)}{ f(t)}(-{A}^2 + {A}^{*2}), \label{36}  \\
g_4(t)&=& f(t)({A}^2 + {A}^{*2}+2A^*A+1). \label{37}
\ea
Now minor arrangements after substituting Eq. (\ref{18}) into the above
equations lead to Eqs. (\ref{24})-(\ref{27}) in the text.

In addition, we see the method for obtaining $\bar{g}_1(t)$ in Eq. (\ref{31}), which is a
coefficient needed for expressing the dynamical phase.
A straightforward evaluation of Eq. (\ref{21}) gives an intermediate formula of the
dynamical phase, which is Eq. (\ref{30}) with Eqs. (\ref{35})-(\ref{37}) and
\be
\bar{g}_1(t)= -\f{1}{ f(t)}({A}^2 + {A}^{*2}-2A^*A-1). \label{38}
\ee
Then, by managing the above equation with the use of Eq. (\ref{18}), we have the eventual
formula of $\bar{g}_1(t)$ which is Eq. (\ref{31}).

\section{Verification that $\Gamma_D(t)$ is a time constant}
To verify that $\Gamma_D(t)$ in Eq. (\ref{D30}) is a time constant, we see its time
derivative. An explicit evaluation with Eq. (\ref{D30}) yields
\ba
\f{d \Gamma_D(t)}{d t} &=& \Bigg[ \omega^2
- \f{1}{f^2(t)}\Bigg( \omega^2 + \f{[\dot{f}(t)]^2}{4} \Bigg)+
\f{\ddot{f}(t)}{2f(t)} \Bigg] \Bigg(A_0^2 \sin [2(\omega T(t) +\theta)]
\nonumber \\
& &-\f{\dot{f}(t)}{4\omega} [1+ 4 A_0^2 \cos^2 (\omega T(t) + \theta)] \Bigg).
\ea
We applied the relation in Eq. (\ref{dtt}) and utilized the fact that
$A_0$ is a time-constant (i.e., $dA_0/dt=0$) in the derivation of this equation.
Now by replacing $\ddot{f}(t)$ in the above equation with Eq. (\ref{dft}),
we readily have ${d \Gamma_D(t)}/{d t}=0$. Hence, $\Gamma_D(t)$ is constant over time.

\section{Geometric, dynamical, and total phases in the Fock states}
The geometric phase, the dynamical phase, and the total phase
in the Fock states with nonstaticity are given by \cite{gow}
\ba
\gamma_{G,n}(t) &=& \left( n+\f{1}{2} \right)
\bigg(\f{c_1+c_2}{2}\omega (t-t_0) - \omega T(t)\bigg)
+\gamma_{G,n}(t_0), \label{gGnt}
\\
\gamma_{D,n}(t) &=& -\left( n+\f{1}{2} \right)\f{c_1+c_2}{2}\omega (t-t_0)
+\gamma_{D,n}(t_0) , \label{gDnt}
\\
\gamma_n(t) &=& - \bigg( n+ \f{1}{2} \bigg)\omega T(t)
+ \gamma_n(t_0), \label{gnt}
\ea
for $t\geq t_0$.
\\
%\\
%{\bf Acknowledgements} \\
%This work was supported by the National Research Foundation of Korea(NRF) grant
%funded by the Korea government(MSIT) (No.: NRF-2021R1F1A1062849).
%\\

%\newpage


\begin{references}

\bibitem{pam} P. A. M. Dirac,
\textquotedblleft Quantised singularities in the electromagnetic field,\textquotedblright~
Proc. Roy. Soc. Lond. A {\bf 133}(821), 60--72 (1931).
DOI: 10.1098/rspa.1931.0130.

\bibitem{pam2} T. V. Mechelen and Z. Jacob,
\textquotedblleft Photonic Dirac monopoles and skyrmions: spin-1 quantization,\textquotedblright~
Opt. Mater. Express {\bf 9}(1), 95--111 (2019).
DOI: 10.1364/OME.9.000095.

\bibitem{ber} M. V. Berry,
\textquotedblleft Quantal phase factors accompanying adiabatic changes,\textquotedblright~
{Proc. R. Soc. Lond. A} {\bf 392}(1802), 45--57 (1984).
DOI: 10.1098/rspa.1984.0023.

\bibitem{ber2} W. Zhao and X. Wang,
\textquotedblleft Berry phase in quantum oscillations of topological materials,\textquotedblright~
Adv. Phys. X, {\bf 7}(1), 2064230 (2022).
DOI: 10.1080/23746149.2022.2064230.

\bibitem{aha} Y. Aharonov and J. Anandan,
\textquotedblleft Phase change during a cyclic quantum evolution,\textquotedblright~
{Phys. Rev. Lett.} {\bf 58}(16), 1593--1596 (1987).
DOI: 10.1103/PhysRevLett.58.1593.

\bibitem{aha2} C. Bengs, M. Sabba, and M. H. Levitt,
\textquotedblleft The Aharonov–Anandan phase and geometric double-quantum excitation
in strongly coupled nuclear spin pairs,\textquotedblright~
J. Chem. Phys. {\bf 158}(12), 124204 (2023).
DOI: 10.1063/5.0138146.

\bibitem{nmu} N. Mukunda and R. Simon,
\textquotedblleft Quantum kinematic approach to the geometric phase. I. General formalism,\textquotedblright~
{Ann. Phys.} {\bf 228}(2), 205--268 (1993).
DOI: 10.1006/aphy.1993.1093.

\bibitem{nmu2} A.-B. A. Mohamed and I. Masmali,
\textquotedblleft Control of the geometric phase in two open qubit–cavity systems
linked by a waveguide,\textquotedblright~
Entropy {\bf 22}(1), 85 (2020).
DOI: 10.3390/e22010085.

\bibitem{nmu3} L. Garza-Soto, N. Hagen, D. Lopez-Mago, and Y. Otani,
\textquotedblleft Differences between the geometric phase and propagation phase:
clarifying the boundedness problem,\textquotedblright~
Appl. Opt. {\bf 63}(3), 645--653 (2024).
DOI: 10.1364/AO.510509.

\bibitem{pol} B. Simon,
\textquotedblleft Holonomy, the quantum adiabatic theorem, and Berry's phase,\textquotedblright~
{Phys. Rev. Lett.} {\bf 51}(24), 2167--2170 (1983).
DOI: 10.1103/PhysRevLett.51.2167.

\bibitem{sfw} K. Y. Bliokh, M. A. Alonso, and M. R. Dennis,
\textquotedblleft Geometric phases in 2D and 3D polarized fields: geometrical,
dynamical, and topological aspects,\textquotedblright~
Rep. Prog. Phys. {\bf 82}(12), 122401 (2019).
DOI 10.1088/1361-6633/ab4415.

\bibitem{mmc} M. Maamache,
\textquotedblleft Ermakov systems, exact solution, and geometrical angles and phases,\textquotedblright~
{Phys. Rev. A} {\bf 52}(2), 936--940 (1995).
DOI: 10.1103/PhysRevA.52.936.

\bibitem{aes} J. Zhang, T. H. Kyaw, S. Filipp, L.-C. Kwek, E. Sj\"{o}qvist, and D. Tong,
\textquotedblleft Geometric and holonomic quantum computation,\textquotedblright~
Phys. Rep. {\bf 1027}, 1--53 (2023).
DOI: 10.1016/j.physrep.2023.07.004.

\bibitem{cvg} G. F. Xu and D. M. Tong,
\textquotedblleft Realizing multi-qubit controlled nonadiabatic holonomic gates with
connecting systems,\textquotedblright~
AAPPS Bull. {\bf 32}(1), 13 (2022).
DOI: 10.1007/s43673-022-00043-6.

\bibitem{rpp}
Y.-H. Lee, G. Tan, T. Zhan, Y. Weng, G. Liu, F. Gou, F. Peng, N. V. Tabiryan, S. Gauza, and S.-T. Wu,
\textquotedblleft Recent progress in Pancharatnam-Berry phase optical elements and the applications for
virtual/augmented realities,\textquotedblright~
{Opt. Data Process. Storage} {\bf 3}(1), 79--88 (2017).
DOI: 10.1515/odps-2017-0010.

\bibitem{jke} J. Lee, Y. Kim, K. Choi, J. Hahn, S.-W. Min, and H. Kim,
\textquotedblleft Digital incoherent compressive holography using a geometric phase metalens,\textquotedblright~
Sensors {\bf 21}(16), 5624 (2021).
DOI: 10.3390/s21165624.

\bibitem{nwh} J. R. Choi,
\textquotedblleft On the possible emergence of nonstatic quantum waves in a static environment,\textquotedblright~
{Nonlinear Dyn.} {\bf 103}(3), 2783--2792 (2021).
DOI: 10.1007/s11071-021-06222-8.

\bibitem{gow} J. R. Choi,
\textquotedblleft Effects of light-wave nonstaticity on accompanying geometric-phase evolutions,\textquotedblright~
{Opt. Express} {\bf 29}(22), 35712--35724 (2021).
%arXiv:2107.03622v1 [quant-ph] (2021).
DOI: 10.1364/OE.440512.

\bibitem{pos} J. R. Choi, K. H. Yeon, I. H. Nahm, and S. S. Kim,
\textquotedblleft Do the generalized Fock-state wave functions have some relations
with classical initial condition?\textquotedblright~
{Pramana-J. Phys.} {\bf 73}(5), 821--828 (2009).
DOI: 10.1007/s12043-009-0150-4.

\bibitem{ncs} J. R. Choi,
\textquotedblleft Analysis of light-wave nonstaticity in the coherent state,\textquotedblright~
Sci. Rep. {\bf 11}, 23974 (2021).
%arXiv:2111.06634v1 [quant-ph] (2021).
DOI: 10.1038/s41598-021-03047-8.

\bibitem{ze0} A. Mostafazadeh,
\textquotedblleft Quantum adiabatic approximation, quantum action, and Berry's phase,\textquotedblright~
{Phys. Lett. A} {\bf 232}(6), 395--398 (1997).
DOI: 10.1016/S0375-9601(97)00391-5.

\bibitem{ze1} G. McCaul, A. Pechen, and D. I. Bondar,
\textquotedblleft Entropy non-conservation and boundary conditions for Hamiltonian dynamical systems,\textquotedblright~
{Phys. Rev. E} {\bf 99}(6), 062121 (2019).
%arXiv:1904.03473v1 [cond-mat.stat-mech] (2019).
DOI: 10.1103/PhysRevE.99.062121.

\bibitem{ze2} O. V. Usatenko, J.-P. Provost, and G. Vall$\acute{\rm e}$e,
\textquotedblleft A comparative study of the Hannay's angles associated with
a damped harmonic oscillator and a
generalized harmonic oscillator,\textquotedblright~
{J. Phys. A} {\bf 29}(10), 2607--2610 (1996).
DOI 10.1088/0305-4470/29/10/035.

\bibitem{conn} J. Y. Zeng and Y. A. Lei,
\textquotedblleft Connection between the Berry phase and the Lewis phase,\textquotedblright~
{Phys. Lett. A} {\bf 215}(5-6), 239--244 (1996).
DOI: 10.1016/0375-9601(96)00254-X.

\bibitem{gis0} P. Martin-Dussaud,
\textquotedblleft Searching for coherent states: From origins to quantum gravity,\textquotedblright~
{Quantum} {\bf 5}, 390 (2021).
DOI: 10.22331/q-2021-01-28-390.

\bibitem{gis} R. J. Glauber,
\textquotedblleft Coherent and incoherent states of the radiation field,\textquotedblright~
{Phys. Rev.} {\bf 131}(6), 2766--2788 (1963).
DOI: 10.1103/PhysRev.131.2766.

\bibitem{opn} W. Schleich and J. A. Wheeler,
\textquotedblleft Oscillations in photon distribution of squeezed states and interference in phase space,\textquotedblright~
Nature {\bf 326}(6113), 574--577 (1987).
DOI: 10.1038/326574a0.

\bibitem{sns} R. Lynch,
\textquotedblleft Simultaneous fourth-order squeezing of both quadrature components,\textquotedblright~
Phys. Rev. A {\bf 49}(4), 2800--2805 (1994).
DOI: 10.1103/PhysRevA.49.2800.

\bibitem{fls} Y. Wang, H. Ye, Z. Yu, Y. Liu, and W. Xu,
\textquotedblleft Sub-Poissonian photon statistics in quantum dot-metal nanoparticles
hybrid system with gain media,\textquotedblright~
Sci. Rep. {\bf 9}, 10088 (2019).
DOI: 10.1038/s41598-019-46576-z.

\bibitem{lou} W. H. Louisell, {\it Quantum Statistical Properties of Radiation}
(John Wiley and Sons, New York, 1973), p. 105.

\bibitem{hka0} N. E. Gürb\"{u}z,
\textquotedblleft Three geometric phases with the visco-Da Rios equation for the hybrid
frame in $R_1^3$,\textquotedblright~
Optik {\bf 248}, 168116 (2021).
DOI: 10.1016/j.ijleo.2021.168116.

\bibitem{hka} H. Kuratsuji and S. Iida,
\textquotedblleft Effective action for adiabatic process: dynamical meaning of Berry and Simon's phase,\textquotedblright~
{Prog. Theor. Phys.} {\bf 74}(3), 439--445 (1985).
DOI: 10.1143/PTP.74.439.

\bibitem{hka1} L. Koens and E. Lauga,
\textquotedblleft Geometric phase methods with Stokes theorem for a general viscous
swimmer,\textquotedblright~
J. Fluid Mech. {\bf 916}, A17 (2021).
DOI: 10.1017/jfm.2021.181.

\bibitem{hka2} P. C. Deshmukh, S. Ghosh, U. Kumar, C. Hareesh, and G. Aravind,
\textquotedblleft A primer on path integrals, Aharonov–Bohm effect and the geometric
phase,\textquotedblright~
Phys. Educ. {\bf 4}(1), 2250005 (2022).
DOI: 10.1142/S2661339522500056.

\bibitem{snb} S. N. Biswas and S. K. Soni,
\textquotedblleft Berry's phase for coherent states and canonical transformation,\textquotedblright~
{Phys. Rev. A} {\bf 43}(10), 5717--5719 (1991).
DOI: 10.1103/PhysRevA.43.5717.

\bibitem{snb2} J. Grain and V. Vennin,
\textquotedblleft Canonical transformations and squeezing formalism in cosmology,\textquotedblright~
J. Cosmol. Astropart. Phys. {\bf 2020}(2), 022 (2020).
DOI 10.1088/1475-7516/2020/02/022.

\bibitem{5} S. Chaturvedi, M. S. Sriram, and V. Srinivasan,
\textquotedblleft Berry's phase for coherent states,\textquotedblright~
{J. Phys. A: Math. Gen.} {\bf 20}(16), L1071--L1075 (1987).
DOI 10.1088/0305-4470/20/16/007.

\bibitem{igp} J. R. Choi,
\textquotedblleft Quadrature squeezing and geometric-phase oscillations in nano-optics,\textquotedblright~
{Nanomaterials} {\bf 10}(7), 1391 (2020).
DOI: 10.3390/nano10071391.

\bibitem{cox1} W. Ding and Z. Wang,
\textquotedblleft ‘Classical’ coherent state generated by curved surface,\textquotedblright~
New J. Phys. {\bf 24}, 113002 (2022).
DOI 10.1088/1367-2630/ac9a9e.

\bibitem{cox2} E. Mungu\'{\i}a-Gonz\'{a}lez, S. Rego, and J. K. Freericks,
\textquotedblleft Making squeezed-coherent states concrete by determining their
wavefunction,\textquotedblright~
Am. J. Phys. {\bf 89}(9), 885--896 (2021).
DOI: 10.1119/10.0004872.

\bibitem{nco1} J. Zhang, T. H. Kyaw, S. Filipp, L.-C. Kwek, E. Sj\"{o}qvist, and D. Tong,
\textquotedblleft Geometric and holonomic quantum computation,\textquotedblright~
Phys. Rep. {\bf 1027}, 1--53 (2023).
DOI: 10.1016/j.physrep.2023.07.004.

\bibitem{nco2} C. P. Jisha, S. Nolte, and A. Alberucci,
\textquotedblleft Geometric phase in Optics: from wavefront manipulation to waveguiding,\textquotedblright~
Laser Photonics Rev. {\bf 15}(10), 2100003 (2021).
DOI: 10.1002/lpor.202100003.

\bibitem{nco3} I. L. Paiva, R. Lenny, and E. Cohen,
\textquotedblleft Geometric phases and the Sagnac effect: foundational aspects and sensing
applications,\textquotedblright~
Adv. Quantum Technol. {\bf 5}(2), 2100121 (2022).
DOI: 10.1002/qute.202100121.

\end{references}
\end{document}